\documentclass[twocolumn]{aastex631}

\shorttitle{Complex Kinematics in the AS 209 Disk}

\graphicspath{{./}{figures/}}

\begin{document}

\title{Molecules with ALMA at Planet-forming Scales (MAPS). Complex Kinematics in the AS 209 Disk Induced by a Forming Planet and Disk Winds}

\correspondingauthor{Maria Galloway-Sprietsma}
\email{mgallowayspriets@ufl.edu}

\author[0000-0002-5503-5476]{Maria Galloway-Sprietsma}
\affiliation{Department of Astronomy, University of Florida, Gainesville, FL 32611, USA}

\author[0000-0001-7258-770X]{Jaehan Bae}
\affiliation{Department of Astronomy, University of Florida, Gainesville, FL 32611, USA}

\author[0000-0003-1534-5186]{Richard Teague}
\affiliation{Department of Earth, Atmospheric, and Planetary Sciences, Massachusetts Institute of Technology, Cambridge, MA 02139, USA}

\author[0000-0002-7695-7605]{Myriam Benisty}
\affiliation{Laboratoire Lagrange, Université Côte d’Azur, CNRS, Observatoire de la Côte d’Azur, 06304 Nice, France} 
\affiliation{Institut de Plan\'etologie et d'Astrophysique de Grenoble (IPAG), Universit\'e Grenoble Alpes, CNRS, F-38000 Grenoble, France}

\author[0000-0003-4689-2684]{Stefano Facchini}
\affiliation{Universit\'a degli Studi di Milano, via Celoria
16, 20133 Milano, Italy}

\author[0000-0003-3283-6884]{Yuri Aikawa}
\affiliation{Department of Astronomy, Graduate School of Science, The University of Tokyo, Tokyo 113-0033, Japan}

\author[0000-0002-2692-7862]{Felipe Alarc\'on}
\affiliation{Department of Astronomy, University of Michigan, 323 West Hall, 1085 S. University Avenue, Ann Arbor, MI 48109, USA}

\author[0000-0003-2253-2270]{Sean M. Andrews} 
\affiliation{Center for Astrophysics \textbar\ Harvard \& Smithsonian, 60 Garden St., Cambridge, MA 02138, USA}

\author[0000-0003-4179-6394]{Edwin Bergin}
\affiliation{Department of Astronomy, University of Michigan, 323 West Hall, 1085 S. University Avenue, Ann Arbor, MI 48109, USA}

\author[0000-0002-2700-9676]{Gianni Cataldi}
\affiliation{National Astronomical Observatory of Japan, 2-21-1 Osawa, Mitaka, Tokyo 181-8588, Japan}

\author[0000-0003-2076-8001]{L. Ilsedore Cleeves}
\affil{Department of Astronomy, University of Virginia, Charlottesville, VA 22904, USA}

\author[0000-0002-1483-8811]{Ian Czekala}
\affiliation{Department of Astronomy and Astrophysics, 525 Davey Laboratory, The Pennsylvania State University, University Park, PA 16802, USA}
\affiliation{Center for Exoplanets and Habitable Worlds, 525 Davey Laboratory, The Pennsylvania State University, University Park, PA 16802, USA}
\affiliation{Center for Astrostatistics, 525 Davey Laboratory, The Pennsylvania State University, University Park, PA 16802, USA}
\affiliation{Institute for Computational \& Data Sciences, The Pennsylvania State University, University Park, PA 16802, USA}

\author[0000-0003-4784-3040]{Viviana V. Guzm\'an}
\affiliation{Instituto de Astrof\'isica, Pontificia Universidad Cat\'olica de Chile, Av. Vicu\~na Mackenna 4860, 7820436 Macul, Santiago, Chile}
\affiliation{N\'ucleo Milenio de Formaci\'on Planetaria (NPF), Chile}

\author[0000-0001-6947-6072]{Jane Huang}
\altaffiliation{NASA Hubble Fellowship Program Sagan Fellow}
\affiliation{Department of Astronomy, University of Michigan, 323 West Hall, 1085 S. University Avenue, Ann Arbor, MI 48109, USA}

\author[0000-0003-1413-1776]{Charles J. Law}\affiliation{Center for Astrophysics \textbar\ Harvard \& Smithsonian, 60 Garden St., Cambridge, MA 02138, USA}

\author[0000-0003-1837-3772]{Romane Le Gal}
\affiliation{Institut de Plan\'etologie et d'Astrophysique de Grenoble (IPAG), Universit\'e Grenoble Alpes, CNRS, F-38000 Grenoble, France}
\affiliation{Institut de Radioastronomie Millim\'etrique (IRAM), 300 rue de la piscine, F-38406 Saint-Martin d'H\`{e}res, France}

\author[0000-0002-7616-666X]{Yao Liu} \affiliation{Purple Mountain Observatory \& Key Laboratory for Radio Astronomy, Chinese Academy of Sciences, Nanjing 210023, China}

\author[0000-0002-7607-719X]{Feng Long}
\altaffiliation{NASA Hubble Fellowship Program Sagan Fellow}
\affiliation{Lunar and Planetary Laboratory, University of Arizona, Tucson, AZ 85721, USA}

\author[0000-0002-1637-7393]{Fran\c cois M\'enard}
\affiliation{Univ. Grenoble Alpes, CNRS, IPAG, 38000 Grenoble, France}

\author[0000-0001-8798-1347]{Karin I. \"Oberg}
\affiliation{Center for Astrophysics \textbar\ Harvard \& Smithsonian, 60 Garden St., Cambridge, MA 02138, USA}

\author[0000-0001-6078-786X]{Catherine Walsh}
\affiliation{School of Physics and Astronomy, University of Leeds, Leeds, UK, LS2 9JT}

\author[0000-0003-1526-7587]{David J. Wilner}\affiliation{Center for Astrophysics \textbar\ Harvard \& Smithsonian, 60 Garden St., Cambridge, MA 02138, USA}

\begin{abstract}

We study the kinematics of the AS~209 disk using the $J=2-1$ transitions of $^{12}$CO, $^{13}$CO, and C$^{18}$O. We derive the radial, azimuthal, and vertical velocity of the gas, taking into account the lowered emission surface near the annular gap at $\simeq1\farcs7$ ($200$~au) within which a candidate circumplanetary disk-hosting planet has been reported previously. In $^{12}$CO and $^{13}$CO, we find a coherent upward flow arising from the gap. The upward gas flow is as fast as $150~{\rm m~s}^{-1}$ in the regions traced by $^{12}$CO emission, which corresponds to about $50\%$ of the local sound speed or $6\%$ of the local Keplerian speed. Such an upward gas flow is difficult to reconcile with an embedded planet alone. Instead, we propose that magnetically driven winds via ambipolar diffusion are triggered by the low gas density within the planet-carved gap, dominating the kinematics of the gap region. We estimate the ambipolar Elsasser number, Am, using the HCO$^+$ column density as a proxy for ion density and find that Am is $\sim 0.1$ at the radial location of the upward flow. This value is broadly consistent with the value at which numerical simulations find ambipolar diffusion drives strong winds. We hypothesize the activation of magnetically-driven winds in a planet-carved gap can control the growth of the embedded planet. We provide a scaling relationship which describes the wind-regulated terminal mass: adopting parameters relevant to 100~au from a solar-mass star, we find the wind-regulated terminal mass is about one Jupiter mass, which may help explain the dearth of directly imaged super-Jovian-mass planets. 

\end{abstract}



\section{Introduction} \label{sec:intro}

Detecting exoplanets during their formation stages allows for a deeper understanding of planet formation processes. However, although there are more than 5000 confirmed exoplanets, only a few of them have been directly detected at a stage when they are still forming \citep{keppler2018,haffert2019,2022NatAs...6..751C}. The Atacama Large Millimeter/submillimeter Array (ALMA) has revolutionized our ability to probe for young, forming planets. ALMA has revealed detailed substructures in continuum emission of protoplanetary disks, such as rings, gaps, and spirals \citep[e.g.,][]{2018ApJ...869L..41A, 2018ApJ...869...17L, 2021MNRAS.501.2934C}. These substructures provide compelling evidence that planets could be present in the disks, although we cannot rule out other origins \citep[see reviews by][]{andrews2020,bae2022_ppvii}.

In addition to continuum observations, by probing the kinematics of the protoplanetary disk gas via molecular line observations, ALMA provides a unique and powerful means to search for young planets. Molecular line observations are capable of discerning subtle localized kinematic perturbations, the so-called velocity kinks, caused by embedded planets \citep{perez_2015ApJ...811L...5P, pinte2018,pinte_2019NatAs...3.1109P, 2020ApJ...890L...9P}. With observations of this nature, one can constrain the surface of the disk in different molecular tracers and therefore understand the three-dimensional velocity structure of the disk. This method is particularly powerful because one can infer the location and mass of the planet \citep[e.g.,][]{izquierdo2021}. Molecular line observations can also probe global-scale dynamics of the protoplanetary disk gas, such as radial changes of the gas velocity \citep{as209_teague_2018ApJ...868..113T,2019Natur.574..378T} and velocity variations along large-scale spirals \citep{teague2019, 2021ApJS..257...18T, wolfer_2022arXiv220809494W}, which can be related to the perturbations created by yet-unseen planets. When multiple molecular lines probing different heights in a disk are used together, one can also probe coherent flows from the surface to the midplane \citep[e.g.,][]{Yu_2021,teague2022}. In addition, circumplanetary disks (CPDs) can be detected with molecular lines, providing unique and strong constraints on their physical and kinematic properties \citep{bae2022}.

Here, we study the kinematics of the AS~209 protoplanetary disk using the $J=2-1$ transitions of $^{12}$CO, $^{13}$CO, and C$^{18}$O obtained as part of the ALMA Large Program Molecules with ALMA at Planet-forming Scales (MAPS; 2018.1.01055.L; \citealt{MAPS_oberg_2021ApJS..257....1O}). AS~209 is a 1--2 Myr-old T Tauri star \citep{2009ApJ...700.1502A, 2018ApJ...869L..41A} and is located 121~pc away in the Ophiuchus star-forming region \citep{gaia_2021}. Previous continuum observations revealed multiple sets of concentric rings and gaps that extend out to $\sim$140 au \citep{guzman_2018ApJ...869L..48G,huang_2018ApJ...869L..42H,sierra2021}, which are theorized to be caused by one or multiple giant planets \citep{fedele_2018A&A...610A..24F,zhang_2018ApJ...869L..47Z}. Molecular line observations also revealed rich annular substructures \citep{huang_2016ApJ...823L..18H, as209_teague_2018ApJ...868..113T, law2021a}. In particular, \citet{as209_teague_2018ApJ...868..113T} {\it kinematically} identified a pressure minimum at $\sim1\farcs9$ (230~au) in $^{12}$CO, which was identified and spatially resolved previously by \cite{guzman_2018ApJ...869L..48G}. The previous work by \cite{as209_teague_2018ApJ...868..113T} used the $^{12}$CO $J=2-1$ transition to measure the rotational velocity of the AS~209 disk and found deviations from Keplerian rotation. More recently, \cite{bae2022} reported a CPD candidate detected in $^{13}$CO $J=2-1$ emission, at the radial separation of $1\farcs7$ (200~au) from the star. With these gas substructures, along with a young, forming planet candidate in the disk, the AS 209 disk warrants a detailed study of its kinematics. 

In this paper, we decompose the line-of-sight velocity into three orthogonal velocity components, namely radial, rotational (or azimuthal), and vertical velocities, for three CO isotopologues, $^{12}$CO, $^{13}$CO, and C$^{18}$O $J=2-1$. As we will show, this allows us to have a more complete three-dimensional view of the kinematic structure of the disk.

This paper is organized as follows. We outline the observations in Section \ref{sec:observations}. In Section \ref{sec:analysis}, we describe the analysis of the data, including the emission surfaces and the velocity profiles, and present the results. In Section \ref{sec:discussion}, we discuss the results focusing on the origin of the velocity structure in the AS~209 disk and its implications. We summarize our findings and discuss future directions in Section \ref{sec:summary}. 

\begin{figure*}
    \centering
    \includegraphics[width=0.9\textwidth]{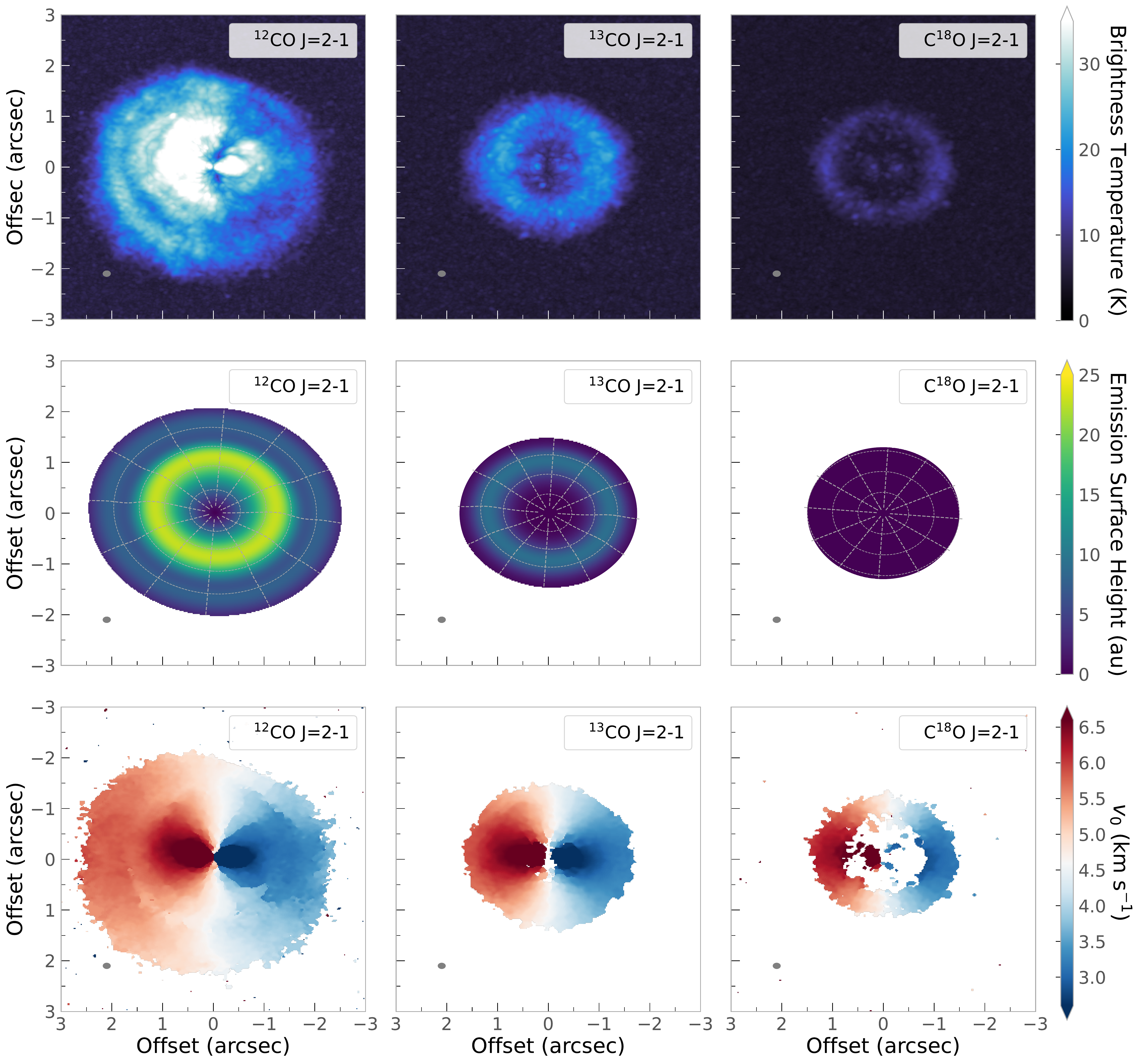}
    \caption{(Upper panels) Peak brightness temperature maps of (left) $^{12}$CO, (middle) $^{13}$CO, and (right) C$^{18}$O $J=2-1$ lines. The $^{12}$CO emission from the western side of the disk experiences foreground cloud contamination as previously reported in independent datasets \citep{oberg_2011ApJ...734...98O, huang_2016ApJ...823L..18H,guzman_2018ApJ...869L..48G,as209_teague_2018ApJ...868..113T, law2021a}. (Middle panels) Emission surface heights above the midplane, taking into account the lowered emission surface across the gap (Model 2; see Section \ref{sec:emission}). For C$^{18}$O, our model is consistent with a flat surface located at the disk midplane. Dashed ellipses and lines show constant radius and azimuth in the disk frame, with intervals of $0\farcs5$ and 30$^\circ$, respectively. (Lower panels) Centroid velocity maps $v_0$ (see Section \ref{sec:velocity_profiles}). Synthesized beams are shown in the lower left corner of each panel.}
    \label{fig:surfaceheightMAPS}
\end{figure*}

\section{Observations} 
\label{sec:observations}

All data used in this work were obtained as part of the ALMA Large Program MAPS\footnote{Data used for this project can be downloaded at the MAPS webpage: \url{https://alma-maps.info/}.}. For the observational setup and calibration process, we refer readers to \cite{MAPS_oberg_2021ApJS..257....1O}. The imaging process is described in \cite{czekala_2021ApJS..257....2C}. As part of the MAPS data release, all images have been post-processed using the \cite{1995AJ....110.2037J} (JvM) correction. For all analysis in this work, we use the \verb|robust = 0.5| weighted, JvM corrected images\footnote{We repeated the analysis using data cubes with a $0\farcs15$ taper and confirmed that the inferred emission surfaces and velocity profiles presented in Section \ref{sec:analysis} do not change significantly. Likewise, we obtain consistent results with JvM-uncorrected cubes as we show in Appendix \ref{sec:no_jvm}.}. The synthesized beam size is 134 mas $\times$ 100 mas for $^{12}$CO $J=2-1$ with a PA of 90.83$^{\circ}$, 140 mas $\times$ 104 mas for $^{13}$CO  $J=2-1$ with a PA of 90.44$^{\circ}$, and 141 mas $\times$ 105 mas for C$^{18}$O $J=2-1$ with a PA of 91.37$^{\circ}$. The rms noise measured in a line-free channel is 0.562 mJy beam$^{-1}$, 0.471 mJy beam$^{-1}$, and 0.339 mJy beam$^{-1}$ for each data cube, respectively. The data were imaged with a channel spacing 200~m~s$^{-1}$, set by the MAPS Program.

In the upper panels of Figure \ref{fig:surfaceheightMAPS}, we present peak brightness temperature maps for $^{12}$CO, $^{13}$CO, and C$^{18}$O $J=2-1$ lines calculated using \verb|bettermoments| \citep{2018RNAAS...2..173T}. The $^{12}$CO $J=2-1$ peak brightness temperature map clearly shows the annular gap at about $1\farcs7$ ($\simeq 200$~au), which is the main feature we focus on in this paper. Additionally, the AS 209 disk suffers from foreground cloud contamination on the western side of the disk, visible in the $^{12}$CO brightness temperature map. \cite{as209_teague_2018ApJ...868..113T} estimated that the cloud absorbs $\sim$30\% of the $^{12}$CO emission along the western side of the disk and showed that this level of perturbations do not impact the kinematic analyses (see their Appendix A.2).

\section{Analysis and Results} \label{sec:analysis}

\subsection{Emission Surface and Disk Geometric Properties}\label{sec:emission}
To begin the characterization of disk kinematics, we first constrain the emission surface for $^{12}$CO, $^{13}$CO, and C$^{18}$O. For our base model, we adopt a power-law emission surface with an exponential taper, given by 
\begin{equation}\label{eqn:emission_surface}
    z(r) = z_0 \, \left( \frac{r}{1^{\prime\prime}} \right)^{\psi}  \exp \left( -\left[ \frac{r}{r_{\rm t}} \right]^{q_{\rm t}} \right),
\end{equation}
where $z(r)$ describes the height $z$ at a given radius $r$, $\psi$ is the power-law exponent, $r_t$ is the characteristic radius for the exponential taper, and $q_t$ is the exponent of the taper term, following \cite{Law_2021,Law_2022}. We note that \cite{Law_2021} already inferred the $^{12}$CO and $^{13}$CO emission surfaces of the AS~209 disk using the same dataset as the one we use in this paper. However, \cite{Law_2021} limited the outer bound of the fit to $1\farcs98$ for $^{12}$CO, which do not cover the full radial extent of the $^{12}$CO disk ($\simeq2\farcs5$), and to $1\farcs35$ for $^{13}$CO, which do not cover the gap around the CPD at $\simeq 1\farcs7$. Because the main goal of this study is to study the kinematics within and around the gap, we opt to fit the emission surfaces adopting larger outer bounds of $2\farcs5$ in $^{12}$CO, $2\farcs0$ in $^{13}$CO, and $1\farcs6$ in C$^{18}$O.

\begin{figure*}
    \centering
    \includegraphics[width=1\linewidth]{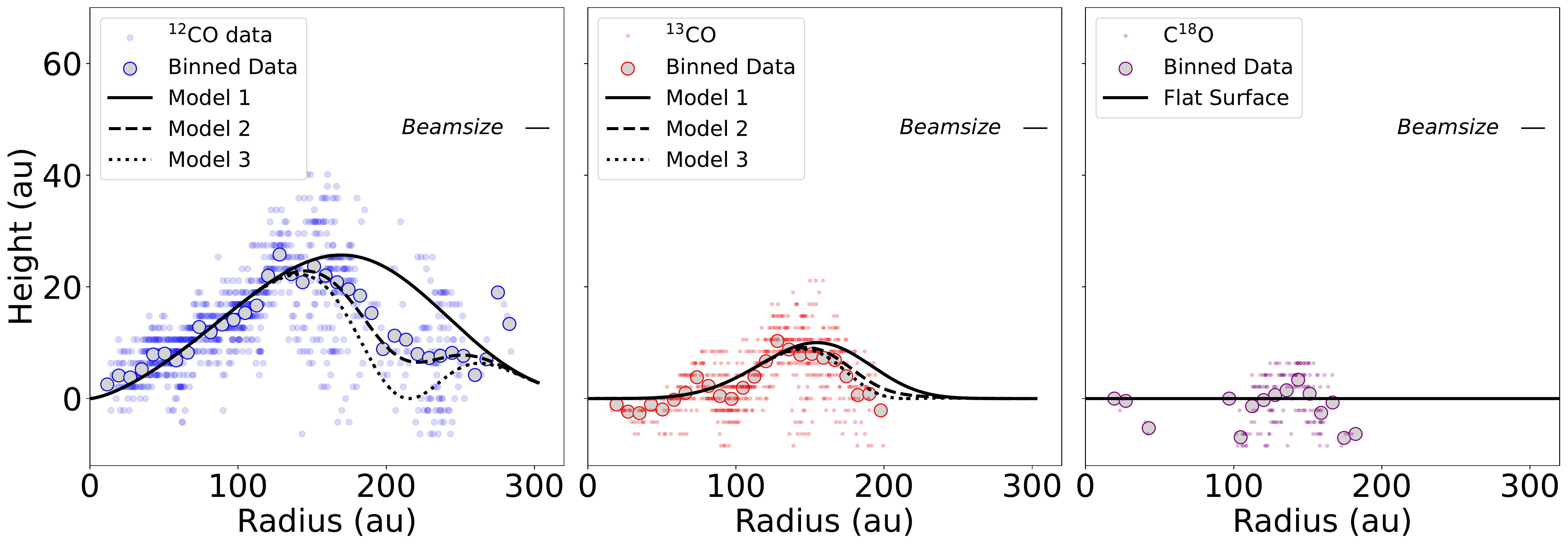}
    \caption{Color dots represent individual pixels inferred by \texttt{disksurf}. Gray filled circles present radially binned pixels, binned by $\sim$0.45 of the beam. Note that the radially binned emission surfaces are shown for visualization purposes only, and the emission surface is fit with individual pixels. We explore three surface models for $^{12}$CO and $^{13}$CO (see Section \ref{sec:emission}): Model 1 (smooth surface, solid curves) assumes a tapered power-law described by Equation (\ref{eqn:emission_surface}), Model 2 (Gaussian gap, dashed curves) adds a Gaussian gap to Model 1, and Model 3 (midplane gap, dotted curves) adopts the same gap center and width as in Model 2, but the gap extends to the midplane (i.e., $A_{\rm gap} = 1$ in Equation (\ref{eqn:emission_surface_gaus})). For C$^{18}$O, we assume a flat surface. The semi-major axis of the synthesized beam is shown in the upper right corner of each panel. Points above one beamsize from the Model 1 curve have been removed for fitting the gap parameters.}
    \label{fig:taperedemissionsurface}
\end{figure*}

To fit the emission surface, we use \verb|disksurf|\footnote{\url{https://disksurf.readthedocs.io/en/latest/}} \citep{disksurf} which implements the method outlined in \cite{2018A&A...609A..47P} who used the asymmetry of the line emission above the disk midplane to infer an emission height. This method allows us to locate emission arising from specific locations in the disk. We then use that information to construct the 3D structure of the emission layer. Following \cite{Law_2021}, we use \texttt{disksurf}'s \texttt{get\_emission\_surface} function to extract the deprojected radius $r$, emission height $z$, surface brightness $I_v$, and channel velocity $v$ for each pixel associated with the emission. We do not exclude channels that suffer from foreground contamination as including the contaminated channels is shown to have no significant effects on the retrieved surface \citep{as209_teague_2018ApJ...868..113T}. For the initial geometric properties used to fit the surface, we assume the disk-center offsets $x_0$ and $y_0$ to be zero, and adopt the position angle PA$=85.8^{\circ}$, inclination $i=35^{\circ}$, and stellar mass $M_*=1.2 M_{\odot}$  from \cite{MAPS_oberg_2021ApJS..257....1O}. We re-fit these parameters later on and confirm that the values we initially adopted describes the data well. For the individual pixels inferred from this procedure, we apply two constraints before we fit the emission surface. First, for all three isotopologues, we implement a minimum $z$ value equal to minus half of the beam semi-major axis. This choice follows the methods from \cite{Law_2021}, where large negative $z/r$ values were removed, but some negative values were allowed to remain to avoid positively biasing the resulting surface. Additionally, for $^{13}$CO and C$^{18}$O, we remove the individual pixels that are above the $^{12}$CO emission surface because $^{13}$CO and C$^{18}$O must be optically thinner than $^{12}$CO. Figure \ref{fig:taperedemissionsurface} shows the individual pixels after data cleaning. We then use the MCMC (Markov Chain Monte Carlo) method from \verb|disksurf| which wraps \texttt{emcee} \citep{2013PASP..125..306F}, adopting 128 walkers, 500 burn-in steps, and 1000 steps to obtain $z_0$, $\psi$, $r_t$, and $q_t$. We confirmed the convergence of the MCMC fit by checking the posterior distribution. Throughout the paper, the emission surface obtained by this process is referred to as Model 1. Table \ref{tab:emission_surf} presents the fitted parameters. 

\begin{table*}[]
\centering
\caption{Emission surface parameters derived for $^{12}$CO and $^{13}$CO $J=2-1$ lines. The errors represent statistical uncertainties and do not account for systematic ones.}
\label{tab:emission_surf}
\begin{tabular}{c|c|c|c|c|c|c|c}
         & z$_0$                   & $\psi$                 & r$_{taper}$                                   & q$_{taper}$            & A$_{gap}$   & r$_{gap}$ & $\sigma_{gap}$ \\
\hline
          & (au)                    &                        & (au)                                          &                        &     & (au)      &     (au)     \\
\hline
$^{12}$CO & 22.99$^{+1.21}_{-1.21}$ & 1.47$^{+0.13}_{-0.11}$ & 217.8$^{+7.26}_{-9.68}$ & 3.69$^{+0.53}_{-0.51}$ & 0.6 & 216.6     & 24.2      \\
$^{13}$CO & 9.68$^{+2.42}_{-2.42}$  & 4.53$^{+0.34}_{-0.58}$ & 152.5$^{+9.68}_{-12.1}$                       & 4.10$^{+0.64}_{-0.92}$ & 0.6 & 216.6     & 24.2     \\
\end{tabular}
\end{table*}

\begin{table*}[]
\centering
\caption{Geometric properties derived for the $^{12}$CO, $^{13}$CO, and C$^{18}$O $J=2-1$ emission assuming a tapered power law with a Gaussian gap (Model 2).The errors represent statistical uncertainties and do not account for systematic ones.}
\label{tab:geometric_prop}

\begin{tabular}{c|c|c|c|c|c}
& x$_0$                   & y$_0$                   & PA                      & M$_*$                    & v$_{LSR}$                \\
\hline
          & (au)                    & (au)                    & ($^{\circ}$)              & (M$_{\odot})$              & (km s$^{-1})$              \\
\hline
$^{12}$CO & $-3.87^{+0.08}_{-0.08}$ & 1.67$^{+0.12}_{-0.12}$  & 84.88$^{+0.07}_{-0.06}$ & 1.24$^{+0.003}_{-0.003}$ & 4.64$^{+0.001}_{-0.001}$ \\
$^{13}$CO & $-1.67^{+0.36}_{-0.36}$ & 0.73$^{+0.29}_{-0.28}$  & 86.01$^{+0.21}_{-0.21}$ & 1.25$^{+0.007}_{-0.008}$ & 4.65$^{+0.003}_{-0.004}$ \\
C$^{18}$O & $-1.57^{+0.76}_{-0.72}$ & -0.57$^{+0.69}_{-0.76}$ & 86.15$^{+0.50}_{-0.51}$ & 1.26$^{+0.01}_{-0.01}$   & 4.66$^{+0.008}_{-0.009}$

\end{tabular}
\end{table*}

Although Equation (\ref{eqn:emission_surface}) describes the overall emission surface well, it cannot describe fine features, such as annular gaps. In particular, the gap at $1\farcs7$ within which a candidate CPD-hosting planet is found \citep{bae2022} cannot be described by Equation (\ref{eqn:emission_surface}). To infer more accurate velocity structures within/around the gap, we add a Gaussian gap to the emission surface obtained in Model 1, adopting the following functional form.
\begin{eqnarray}
\label{eqn:emission_surface_gaus}
    z(r) & = & z_0 \, \left( \frac{r}{1^{\prime\prime}} \right)^{\psi}  \exp \left( -\left[ \frac{r}{r_{\rm t}} \right]^{q_{\rm t}} \right) \nonumber \\
    & \times & \left( 1 - A_{\rm gap} \exp\left[- \frac{(r-r_{\rm gap})^2}{2\sigma_{\rm gap}^2}\right]\right)
\end{eqnarray}
Here, $A_{\rm gap}$, $r_{gap}$, and $\sigma_{\rm gap}$ describe the depth of the gap, radial location of the center of the gap, and radial width of the gap, respectively. After obtaining the tapered-power law parameters using the aforementioned methods, we fit for only the gap parameters using \verb|scipy.optimize.curve_fit|. For this process, we remove individual pixels above one beam from the Model 1 emission surface for a better convergence of the fit. The removed pixels through this procedure is less than 10\% of the entire pixels. We note that removing these individual pixels at high altitude estimates a deeper gap than would otherwise be found if these pixels were included.  However, as we show below, the inferred velocity profiles are insensitive to the depth of the gap. As for Model 1, we sample the posterior distributions using an MCMC approach, adopting 128 walkers, 500 burn-in steps, and 1000 steps. 

From now on, we refer to this surface with a Gaussian gap as Model 2, and this model is the main model we will use for our analysis. We do not fit the gap in $^{13}$CO separately because the $^{13}$CO emission is weak beyond $\simeq1\farcs5$ and does not probe the full extent of the gap. Instead, we adopt the best-fit gap parameters from the $^{12}$CO data. As we found that the C$^{18}$O emission surface is consistent with a flat surface at the disk midplane, we do not introduce a gap in the C$^{18}$O surface (see \citealt{2022arXiv221208667L} for flat C$^{18}$O emission surfaces in other disks). As such, throughout this paper we adopt a single model with zero emission height for C$^{18}$O. Figure \ref{fig:taperedemissionsurface} shows emission surfaces from all the models. 
Table \ref{tab:emission_surf} presents the best-fit gap parameters.

Finally, we allow the Gaussian gap to reach the midplane by setting $A_{\rm gap}=1$, which we denote as Model 3. The purpose of having this hypothetical model is to allow the emission surface to reach the disk midplane and examine the effect of the gap depth in the derived velocity profile.

\begin{figure*}
    \centering
    \includegraphics[width=\textwidth]{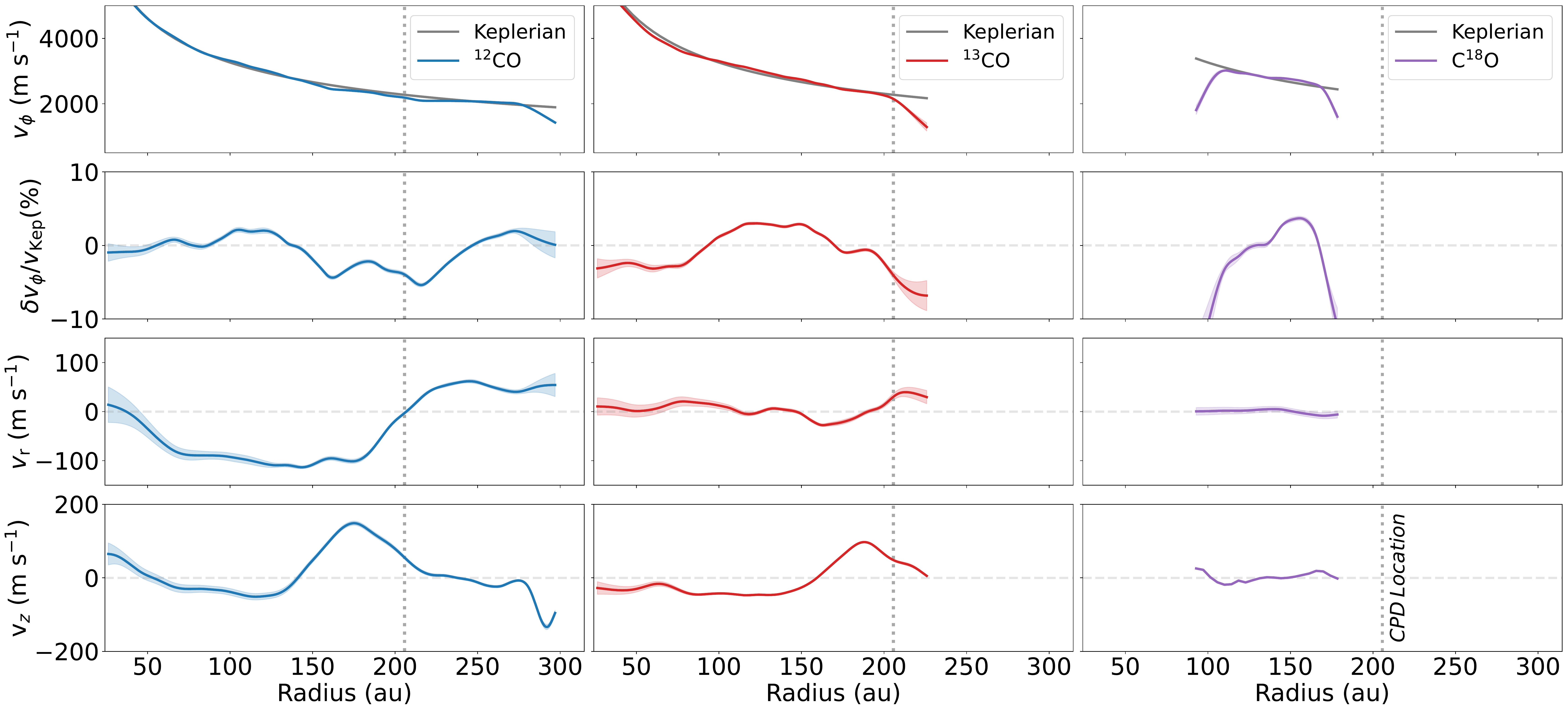}
    \caption{Velocity profiles for $^{12}$CO (left column, blue), $^{13}$CO (middle column, red), and C$^{18}$O (right column, purple). For $^{12}$CO and $^{13}$CO, we adopt the emission surfaces described by a tapered power-law with a Gaussian gap (i.e., Model 2; Equation \ref{eqn:emission_surface_gaus}), while a flat emission surface at the midplane is assumed for C$^{18}$O. For each isotopologue, the first rows show the rotational velocity $v_\phi$, the second rows show the perturbations in $v_\phi$ from Keplerian rotation $v_{\rm Kep}$, $\delta v_{\phi}/v_{\rm Kep} \equiv (v_{\phi}-v_{\rm Kep})/v_{\rm Kep}$, the third rows show the radial velocity $v_r$, and the fourth rows the  vertical velocity, $v_z$, respectively. The vertical dark grey dotted line shows the radial location of the CPD. The shaded regions represent the uncertainties on the velocities; these are statistical uncertainties and do not account for systematic ones.}
    \label{fig:vphi_vrad}
\end{figure*}

Once the emission surfaces are fitted, we take the best-fit values to infer the geometric properties of the disk using \verb|eddy|\footnote{\url{https://eddy.readthedocs.io/en/latest/}} \citep{eddy}. We fit the disk center offset $x_0$ and $y_0$, disk position angle PA, dynamical stellar mass $M_*$, and the LSR velocity of the target $v_{\rm LSR}$, while the disk inclination is fixed to $35^\circ$, a value constrained by high-resolution continuum data \citep{huang_2018ApJ...869L..42H}. We use an MCMC method with the same setup mentioned previously. The geometric properties obtained using the Model 2 emission surface are listed in Table \ref{tab:geometric_prop}, while those derived using Model 1 and Model 3 are listed in Table \ref{tab:geometric_model1} and Table \ref{tab:geometric_model3} in Appendix \ref{sec:app_models}. The geometric properties obtained via this method are broadly consistent with the dust-based values obtained in \cite{huang_2018ApJ...869L..42H}, who finds a position angle of 85.76$\pm$0.16$^{\circ}$. Our position angle values for $^{13}$CO (86.01$^{+0.21}_{-0.21}$$^{\circ}$) and C$^{18}$O (86.148$^{+0.50}_{-0.51}$$^{\circ}$) are closer to the value obtained via continuum fitting by \cite{huang_2018ApJ...869L..42H} likely because they trace closer to the midplane. These geometric properties are also consistent with those from \cite{MAPS_oberg_2021ApJS..257....1O}.

\begin{figure*}
    \centering
    \includegraphics[width=0.9\textwidth]{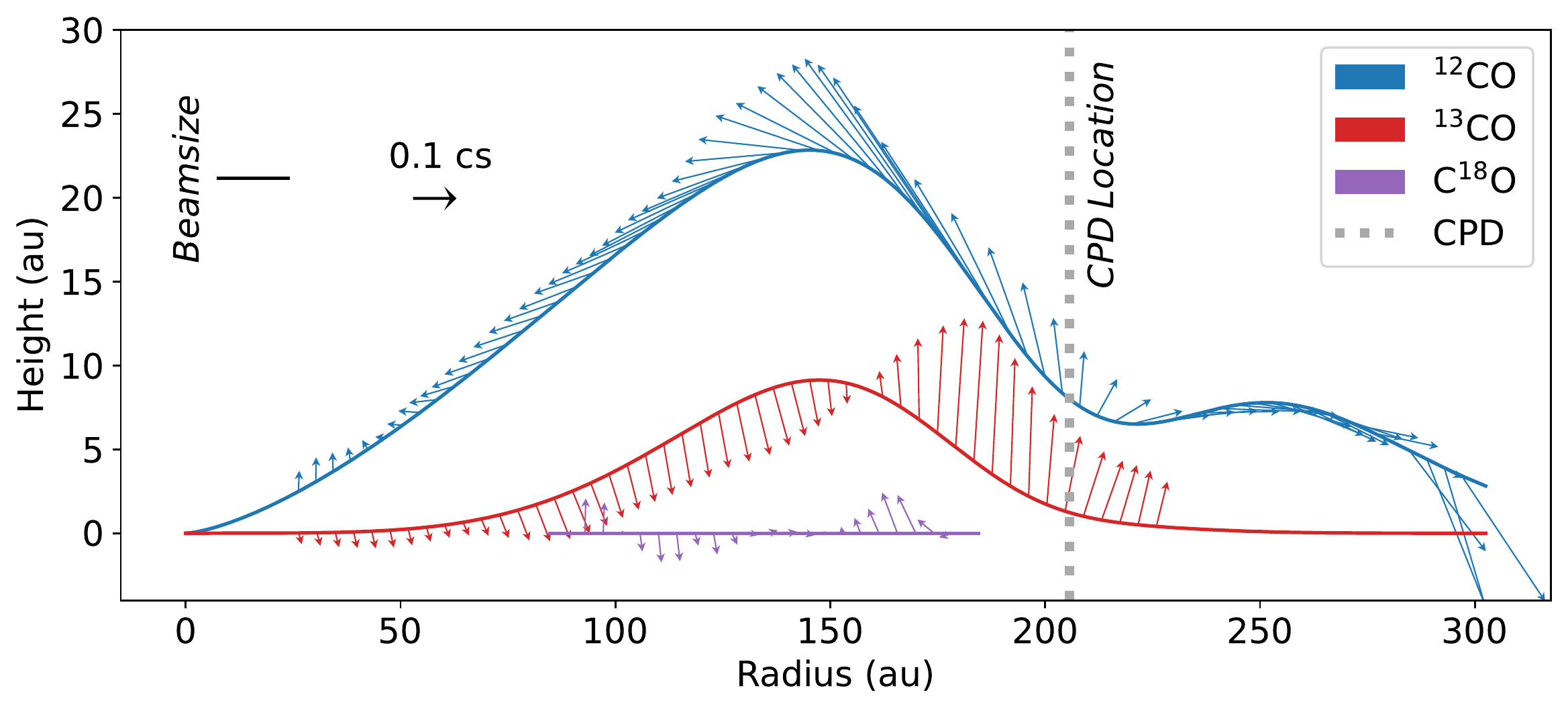}
    \caption{Gas flows in the $r - z$ plane for $^{12}$CO (blue), $^{13}$CO (red), and C$^{18}$O (purple). Note the strong upward gas flows arising from the gap. The emission surfaces are shown with thick curves. The arrows are scaled to the local sound speed $c_s$, where we use the two-dimensional $r-z$ gas temperature distribution inferred by \cite{Law_2021}. The size of the semi-major axis of the synthesized beam ($^{12}$CO) is shown in the upper left corner. An arrow showing $10\%$ of the sound speed is also presented in the upper left corner. The vertical dark grey dotted line shows the location of the CPD.}
    \label{fig:COrz_quiver}
\end{figure*}

\subsection{Velocity Profiles} \label{sec:velocity_profiles}

To infer the velocity profiles, we first make maps of the line centers $v_0$, using the quadratic method from \verb|bettermoments| \citep{bettermoments}. The resulting $v_0$ maps are shown in the bottom panels of Figure \ref{fig:surfaceheightMAPS}. Then, with the derived emission surface and disk geometric properties, we decompose $v_0$ into the radial, rotational, and vertical velocities, following \cite{teague_bae_2018ApJ...860L..12T,as209_teague_2018ApJ...868..113T,2019Natur.574..378T}.

This is done by breaking apart the following equation 
\begin{equation}
\label{eqn:velocity_profile}
    v_{0} = v_{\phi} \cos(\phi) \sin(|i|) + v_{r} \sin(\phi)\sin(i) - v_z \cos(i) + v_{\rm LSR}
\end{equation}
assuming that $v_\phi$ and $v_r$ are azimuthally symmetric, where $v_{\phi}$ is the rotational velocity, $v_{r}$ is the radial velocity, $v_z$ is the vertical velocity, $i$ is the inclination of the disk\footnote{In Equation (\ref{eqn:velocity_profile}), positive $i$ represents a disk that is rotating in a counter-clockwise direction, while negative $i$ describes a clockwise rotation \citep{pinte_review_https://doi.org/10.48550/arxiv.2203.09528}. Because the AS~209 disk rotates clockwise, we adopt $i=-35^\circ$.}, and $\phi$ is the azimuthal angle in the frame of reference of the disk.
In practice, we use the \verb|get_velocity_profile| module from \verb|eddy| \citep{eddy} with 20 iterations to obtain the stacked spectra, each of which use a random sample of independent pixels; a weighted average is then taken over these 20 samples to calculate $v_{r}$ and $v_{\phi}$. We choose this number of iterations based on \cite{Yu_2021}, who found that the gradient of the average standard deviation of the results flattens after about 20 iterations.

As in \cite{as209_teague_2018ApJ...868..113T}, we model the stacked spectrum with a Gaussian Process, which allows for a more flexible and robust model \citep{2017AJ....154..220F}. As can be seen in Equation (\ref{eqn:velocity_profile}), the vertical velocity has no dependence on $\phi$ and is thus not directly calculated by shifting and stacking spectra. Instead, to calculate $v_z$, we exploit Equation (\ref{eqn:velocity_profile}) and subtract projected radial and rotational velocities, along with $v_{\rm LSR}$, from the $v_{0}$ map, following \cite{Yu_2021}. 

The resulting velocity profiles for $^{12}$CO, $^{13}$CO, and C$^{18}$O are shown in Figure \ref{fig:vphi_vrad}.
Looking at the rotational velocity first, we find evidence of super- and sub-Keplerian rotation in $^{12}$CO on the order of $\pm5\%$ of the background Keplerian rotation.
The sub-Keplerian rotation is most significant at $\sim150 - 240$~au and has a double-peaked profile. At $\sim80 - 130$~au and $\gtrsim 230$~au, the disk rotation is super-Keplerian, up to about $2\%$ of the background Keplerian rotation. Overall, the $^{12}$CO rotational velocity profile is consistent with what was previously inferred by \citet{as209_teague_2018ApJ...868..113T}.
The $^{13}$CO emission shows a rotational velocity profile that is broadly consistent with $^{12}$CO: the disk at $\sim90 - 170$~au has super-Keplerian motion.
Additionally, we find a rapid transition to sub-Keplerian rotation beyond $190$~au. We conjecture that 
this is due to lower signal-to-noise ratio (SNR). We find a similar rapid transition to sub-Keplerian rotation in C$^{18}$O beyond $170$~au, likely due to low SNR.

Next, the radial velocity profile in $^{12}$CO shows a change in sign, from about $-100~{\rm m~s}^{-1}$ to $50~{\rm m~s}^{-1}$, around 200~au. There is not a similar trend in $^{13}$CO, and the magnitude of the radial velocity is much smaller than that of $^{12}$CO, within $\pm20~{\rm m~s}^{-1}$. The radial velocity of C$^{18}$O is consistent with zero within uncertainties.

Lastly, we find a large upward vertical velocity flow in $^{12}$CO. This upward vertical motion is persistent from 140 to 220~au and has a maximum velocity of about $150~{\rm m~s}^{-1}$ at a radius of $\simeq 177$~au, which corresponds to about $6\%$ of the local Keplerian speed or $50\%$ of the local sound speed adopting the two-dimensional $r-z$ gas temperature distribution inferred by \cite{Law_2021}. The vertical velocity in $^{13}$CO emission also shows evidence of large coherent upward motions from 160 to 220~au, with a maximum velocity of $85~{\rm m~s}^{-1}$ at 193~au. In C$^{18}$O, we see a much smaller upward motion, reaching a maximum of about $10~{\rm m~s}^{-1}$, but note that C$^{18}$O emission is weak and does not probe the radial regions where strong upward motions are seen in $^{12}$CO or $^{13}$CO. These velocity profiles are broadly consistent with what are found by \cite{Izquierdo2023}, where the authors carried out an independent kinematic analysis on the same data obtained by the MAPS program. In Section \ref{sec:vertical_flow_origin} we discuss the potential origin of these coherent, large-scale upward flows.

We examined how (in)sensitive the inferred velocity profiles are to the assumed emission surface by repeating the analysis and deriving velocity profiles adopting Model 1 (tapered power-law emission surface without a gap) and Model 3 (tapered power-law emission surface with a Gaussian gap that reaches the midplane). As we show in Figure \ref{fig:velocity_profiles_compare} in Appendix \ref{sec:app_models}, varying the emission surfaces does not have a significant impact on the velocity profiles. For the rest of the paper, we thus opt to use Model 2 for our discussion. To help visualize the inferred velocity flows along with the emission surfaces, in Figure \ref{fig:COrz_quiver} we depict the gas flows in the $r-z$ plane. As shown, it is apparent that the large upward motions in $^{12}$CO and $^{13}$CO coincide with the gap in the disk.

\begin{figure*}
    \centering
    \includegraphics[width=1\textwidth]{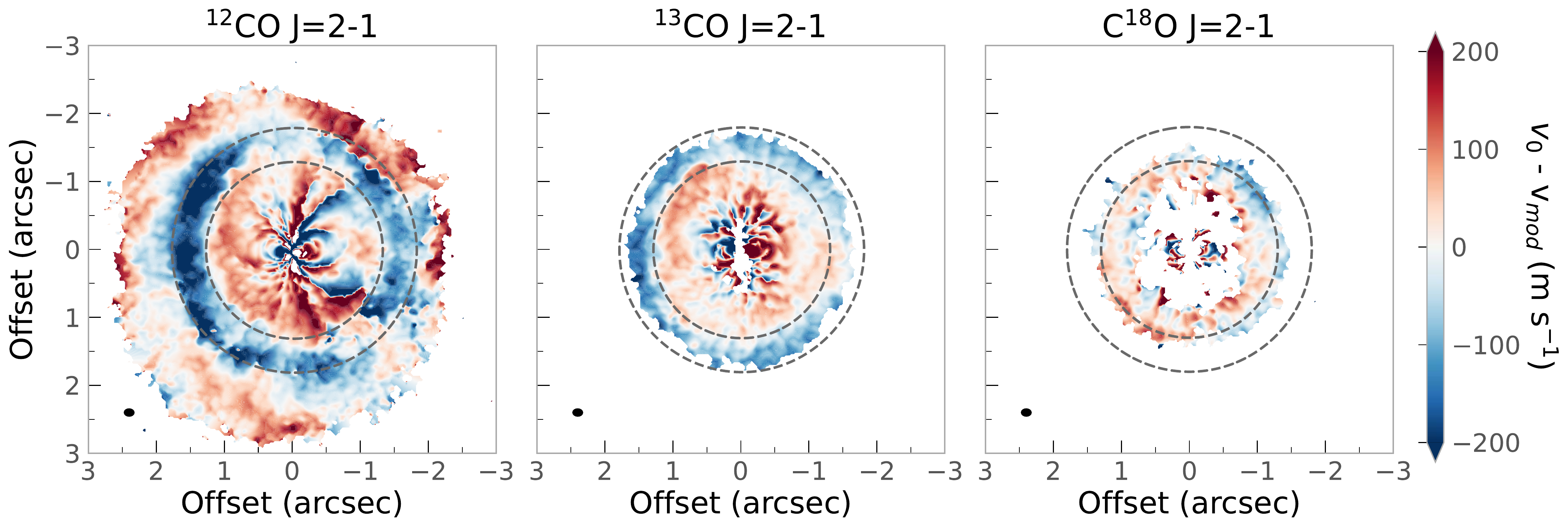}
    \caption{Deprojected velocity residual maps for $^{12}$CO, $^{13}$CO, and C$^{18}$O, calculated by subtracting best-fit Keplerian models ($v_{\rm mod}$) from the centroid velocity maps ($v_0$; see Figure \ref{fig:surfaceheightMAPS} bottom panels). 
    Note that in $^{12}$CO and $^{13}$CO, the velocity structures within the gap at $\simeq 1\farcs5 - 2''$ are present around the entire azimuth. The dashed circles show lines of constant radius at $1\farcs3$ and $1\farcs7$, highlighting the velocity structure associated with upward vertical motions arising from the gap.
    Synthesized beams are shown in the lower-left corner of each panel. }
    \label{fig:residual_v0}
\end{figure*}

In addition to searching for kinematic structures in azimuthally averaged radial profiles of the velocity, we investigate structure within the deprojected residual velocity maps. To do so, we calculate a best-fit Keplerian model with \verb|eddy|, adopting the emission surfaces and the disk geometric properties constrained as in Section \ref{sec:emission}. This produces a model map, $v_{\rm mod}$, which we subtract from the line centroid map, $v_0$ (shown in Figure 1). Figure \ref{fig:residual_v0} shows the resulting residual maps in the deprojected, disk plane for all three CO isotopologues. As seen in $^{12}$CO and $^{13}$CO, the velocity structure in the residual maps is mostly azimuthally symmetric, indicating that the velocity perturbation contributions are largely from the vertical component \citep{teague2019}. We find no clear asymmetric features associated with the planet candidate proposed by \citet{bae2022}. However, as we will discuss in Section \ref{sec:vertical_flow_origin}, the kinematics of the gap is likely dominated by disk winds, not the planet, and we emphasize that lack of asymmetric features in the residual velocity maps does not dispute the presence of a planet.

\section{Discussion} \label{sec:discussion}

\subsection{Origin of the Vertical Flows}
\label{sec:vertical_flow_origin}

The most prominent kinematic structure found in the AS~209 disk is the upward flow in $^{12}$CO and $^{13}$CO, arising from the gap at $1\farcs2 - 1\farcs8$ ($145 - 218$~au). In this section, we explore several possibilities to explain the upward flows.

\subsubsection{Giant Planet}

Giant planets are expected to perturb the velocity structure of the disk. When a giant planet opens a gap, steep density gradients develop on the sides of the gap, driving sub-/super-Keplerian rotation \citep{kanagawa2015} as well as a downward flow back into the midplane \citep{kley2001,gressel2013,2014Icar..232..266M, 2014ApJ...782...65S, fung_2016ApJ...832..105F}. 
However, this is not the trend we see in the AS~209 disk. As shown in Figure \ref{fig:COrz_quiver}, the $^{12}$CO and $^{13}$CO vertical velocity patterns reveal upward motion at the center of the gap which tapers off at the gap edges---a meridional {\it fountain}. This upward motion is seen across the gap over a broad range of azimuth (only a small section of azimuth in the 4 o'clock direction does not exhibit the upward flows within the gap, see Figure \ref{fig:residual_v0}), so it is unlikely that the observed upward flows are associated with a jet or outflow arising locally from the embedded planet. 


Alternatively, one might ask if we are seeing downward flows toward the midplane from the back side of the disk. This may be possible when the front side of the disk is sufficiently optically thin; however, $^{12}$CO remains optically thick within the gap, supported by the fact that the $^{12}$CO brightness temperature within the gap is ${>}20$ K \citep{Law_2021} and that the CPD is visible only in $^{13}$CO and not in $^{12}$CO \citep{bae2022}. Overall, the upward flows seen in AS 209 are not straightforward to reconcile with the presence of a giant planet alone. 

\subsubsection{Ambipolar Diffusion-Driven Winds} \label{sec:amb_driv_wind}

To explain both the presence of the CPD-hosting planet previously reported in \citet{bae2022} and the azimuthally symmetric upward gas flows found in this paper, we propose a scenario where the low density within the planet-carved gap triggers magnetically driven winds via ambipolar diffusion. Ambipolar diffusion is the dominant non-ideal magnetohydrodynamic (MHD) effect when the underlying gas density and ionization levels are low \citep{wardle2007}. When ambipolar diffusion dominates the gas dynamics, ions that are coupled to the magnetic fields can drag neutral molecules/atoms, driving winds \citep[][see also the review by \citealt{lesur2022}]{bai2013,gressel2015,bethune2017,suriano2018,2022_hu_MNRAS.516.2006H}.

To examine this possibility more quantitatively, we estimate the ratio of the ion-neutral drift time to the dynamical time by calculating the ambipolar Elsasser number Am, given by 
\begin{equation}
    {\rm Am} \equiv \frac{v_A^2}{\eta_A \Omega_K},
\label{eq:Am}
\end{equation}
where $v_A$ is the Alfv\'en velocity, $\eta_A$ is the ambipolar diffusivity, and $\Omega_K$ is the Keplerian frequency \citep{bai2011}. Using $v_A  = B/\sqrt{4\pi\rho}$ and $\eta_A = B^2/(4\pi\gamma\rho\rho_i)$, where $B$ is the magnetic field strength, $\rho$ is the density of the neutral gas, and $\rho_i$ is the density of the ionized gas, Equation (\ref{eq:Am}) turns into 
\begin{equation}
    {\rm Am} = \frac{\gamma \rho_i}{\Omega_K}.
\label{eq:Am_diffusion}
\end{equation}
Here, $\gamma\equiv \langle \sigma v \rangle_i/(m_n + m_i)$ where $\langle\sigma v\rangle_i$ is the momentum transfer rate coefficient for an ion-neutral collision, given by 
\begin{equation}
    \langle\sigma v\rangle_i = 2.0 \times 10^{-9} \left( \frac{m_H}{\mu} \right)^{1/2} {\rm {\rm cm}^3~{\rm s}^{-1}},
\end{equation}
where $m_n$ and $m_i$ are the mass of neutral and ion, and $\mu \equiv m_n m_i /(m_n+m_i)$ is the reduced mass \citep{draine2011}. 

\begin{figure}
    \includegraphics[width=85mm]{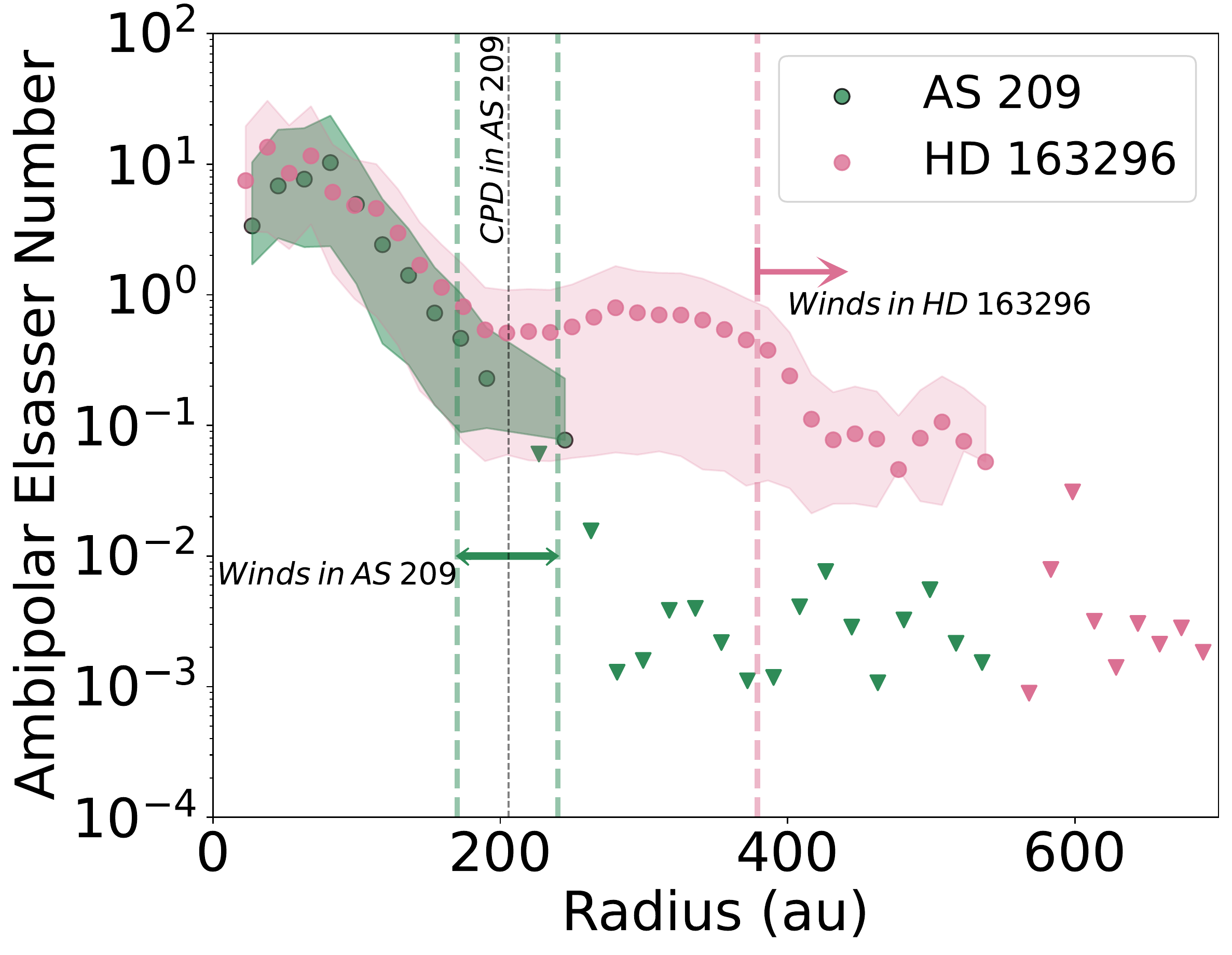}
    \caption{Ambipolar Elsasser number, Am, calculated for the AS 209 disk (green) and the HD 163296 disk (pink). The triangles show upper limits on Am when HCO$^+$ $J=1-0$ is not detected. The shaded regions represent the uncertainty on the measurements, which are derived from the uncertainty in the $N$(HCO$^+$) measurement. The green dashed lines and arrows show the radial region where winds are found in AS 209, coincident with the annular gap in $^{12}$CO and the CPD. The grey dashed line shows the radial location of the CPD in AS 209 \citep{bae2022}. The pink dashed line and arrow show the radial region where winds are found in HD 163296 \citep{2019Natur.574..378T}. }
    \label{fig:ambipolar}
\end{figure}

Thermochemical models of protoplanetary disks suggest that HCO$^+$ is the most abundant molecular ion in the warm molecular layer where CO gas is abundant \citep[e.g.,][]{aikawa2015}. In fact, in all five disks observed in MAPS, the HCO$^+$ column density is greater than that of N$_2$H$^+$ and N$_2$D$^+$ (two other ions that are believed to be abundant in protoplanetary disks) by more than an order of magnitude \citep{2021_Aikawa_ApJS..257...13A}. Assuming that HCO$^+$ and H$_2$ are the dominant ions and neutrals in protoplanetary disks, we obtain $\gamma = 2.82 \times 10^{13}~{\rm cm^3~g^{-1}~s^{-1}}$.  To compute the ion density $\rho_i$, we use the observationally constrained column density of HCO$^+$ by \cite{2021_Aikawa_ApJS..257...13A}\footnote{Available to download at the MAPS webpage: \url{https://alma-maps.info/}.} and divide it by the gas pressure scale height, $\rho_i=m_i n_i = m_{\rm HCO^+} \times N({\rm HCO}^+) / H(r)$, motivated by thermochemical models where HCO$^{+}$ forms a layer having an approximately constant volume density \citep{2021_Aikawa_ApJS..257...13A}. We calculate the scale height using a power-law with a flaring index determined by \cite{zhang2021} for the AS~209 disk.  Figure \ref{fig:ambipolar} shows the derived ambipolar Elsasser number as a function of disk radius. Within the inner $\sim100$~au Am is about 10, but it drops to $\sim0.1$ at 200~au due to the low HCO$^+$ density.
Recent non-ideal MHD simulations have shown that when Am drops to $\sim 1$ ambipolar diffusion starts to quench the MRI \citep{2011_bai_stone_ApJ...736..144B} and launches  winds \citep{bai2013,gressel2015,suriano2018}. The inferred ambipolar Elsasser number of $\sim0.1$ at the radial region of emerging vertical flows is thus broadly consistent with these numerical simulations. 
In case the outer disk is transparent to UV radiation and C$^{+}$ dominates the ion density instead of HCO$^{+}$, our Am estimates would provide a lower limit.

In the HD~163296 disk, \citet{2019Natur.574..378T,teague2022} found upward\footnote{The sign of the vertical velocity extracted in previous papers, \citet{2019Natur.574..378T, 2021ApJS..257...18T}, was incorrect and needs to be flipped.} meridional flows, most prominently at 240~au, the radial location of a kinematically inferred planet \citep{pinte2018,teague_bae_2018ApJ...860L..12T}, and radially outward disk winds beyond $\sim380$~au. Are these findings consistent with the picture we propose for AS~209? In order to answer this question, we compute the ambipolar Elsasser number in the HD~163296 disk using the same methods we applied to the AS~209 disk, using the HCO$^+$ column density from \cite{2021_Aikawa_ApJS..257...13A} and the scale height from \cite{zhang2021}. The resulting radial profile of the ambipolar Elsasser number is shown in Figure \ref{fig:ambipolar}. As in AS~209, the ambipolar Elsasser number is $\sim10$ in the inner $\sim100$~au of the disk. At 240~au in HD 163296, the ambipolar Elsasser number is $\sim0.5$ and beyond $\sim380$~au the ambipolar Elsasser number drops to $\sim0.1$. 
Within the two disks, we find an overall trend that winds appear in the radial regions having a low ambipolar Elsasser number of $\sim0.1 - 0.5$,  suggesting that winds driven by ambipolar diffusion may be common in low-density regions of protoplanetary disks.

Besides a low density, ambipolar diffusion is more efficient in the presence of strong magnetic fields. We estimate the required magnetic field strength in the AS 209 disk by using the magnetic diffusion numbers from \cite{wardle2007} who defines the ambipolar regime as being dominant when 1 $\ll$ $\beta_i$ $\ll$ $\beta_e$, where the magnetic diffusion numbers, $\beta_i$ and $\beta_e$, are given by
\begin{equation} 
\label{eqn:beta_i}
    \beta_i \approx 4.6 \times 10^{-3} \left( {B \over 1~G} \right) \left( n_{H} \over 10^{15}~{\rm cm}^{-3}  \right)^{-1}
\end{equation}
and 
\begin{equation}
\label{eqn:beta_e}
    \beta_e = 3.5 \left( {B \over 1~G} \right) \left( {n_{H} \over 10^{15}~{\rm cm}^{-3}} \right)^{-1} \left( \frac{T}{100K}\right)^{-1/2}.
\end{equation}
In the above equations, $B$ is the magnetic field in Gauss and $n_{H}$ is the number density of hydrogen nuclei in units of ${\rm cm}^{-3}$. A lower limit on the magnetic field strength which satisfies $\beta_i \gg 1$ is obtained using the hydrogen column density. Note that the second part of the inequality, $\beta_i$ $\ll$ $\beta_e$, is always satisfied with gas temperatures of tens to hundreds Kelvin, as can be seen from Equations (\ref{eqn:beta_i}) and (\ref{eqn:beta_e}). We compute the hydrogen column density N(H$_2$) using the gas surface density derived in \citet[][see their Section 4.1 and Figure 16]{zhang2021}. 
At 200~au, N(H$_2$) $\simeq 1.8 \times 10^{21}$ cm$^{-2}$. Adopting the scale height at 200~au of $H=14.3$~au from \citet{zhang2021}, the number density of hydrogen nuclei at the midplane is $n_H = (1/2) \times N(H_2)/\sqrt{2\pi}H \simeq 1.7\times10^6~{\rm cm}^{-3}$. Inserting this hydrogen nuclei number density into Equation (\ref{eqn:beta_i}), we find that a weak magnetic field strength of $B \gg 0.36~\mu$G is sufficient for ambipolar diffusion to dominate at 200~au. Note also that the weak required magnetic field strength is consistent with non-detection of magnetic fields in the AS~209 disk (3$\sigma$ upper limits of a few mG) via observations of Zeeman splitting of the CN $N=2–1$ line \citep{2021_zeeman_mag_ApJ...908..141H}.

\subsubsection{Vertical Shear Instability}

Vertical shear in the rotational velocity of the disk gas can lead to an instability that can produce vertical flows when saturated \citep{nelson2013}. \citet{barraza2021} showed that vertical flows driven by the vertical shear instability (VSI) can manifest as nearly concentric rings of upward and downward flows in the Keplerian-subtracted centroid velocity maps of molecular line emission. However, we conjecture that the VSI is less likely to be the origin of the vertical flows seen in the AS~209 disk because the radial extent of the vertical flow in the AS~209 disk is much larger than what is typically seen in numerical simulations of the VSI. The radial width of the VSI-induced vertical flows in numerical simulations is about a gas scale height \citep{nelson2013,barraza2021}. On the other hand, the upward flow in the AS~209 disk spans about $0\farcs5 \simeq 60.5$~au which corresponds to about 4 scale heights at $1\farcs7 \simeq 206$~au adopting the midplane scale height of $0\farcs12 \simeq 14.7$~au from \citet{zhang2021}. 

In summary, we conclude that ambipolar diffusion-driven winds from a planet-carved gap is the most viable origin for the observed vertical flows in the AS~209 disk.

\subsection{Can winds stop the growth of giant planets?}
 
In the traditional meridional circulation picture without winds, the rate at which a planet would grow depends on the rate of the circumstellar disk gas flowing into the planet-carved gap \citep[e.g.,][]{2014Icar..232..266M}. In this picture, planets can continuously grow in mass until the circumstellar disk loses most of its mass. Indeed, hydrodynamic simulations showed that the mass-doubling time for a Jovian-mass planet is of order of $100 - 1000$ orbital times, which can be much shorter than the lifetime of protoplanetary disks depending on the radial location of the planet \citep[e.g.,][]{kley1999,lubow1999}. 

In our modified picture considering ambipolar diffusion-driven winds, the mass outflow rate via winds can exceed the mass inflow rate toward the gap, in which case the growth of the embedded planet can be limited or even ceased. In the AS~209 disk, we can estimate the mass loss rate from the annular gap using the following equation:
\begin{equation}
    \dot{M}_{\rm wind} = \int_{r_{\rm in}}^{r_{\rm out}}{2\pi r \rho v_{\rm wind} {\rm d}r},
\end{equation}
where $r_{\rm in}=1\farcs3$ ($\sim$160 au) and $r_{\rm out}=2\arcsec$ ($\sim$240au) are the inner and outer boundaries of the wind-launching region, and $\rho$ and $v_{\rm wind}$ are the gas density and speed of the wind. For simplicity, we opt to use the midplane density, $\rho = \rho_{\rm mid}$, and the vertical velocity of $^{12}$CO, $v_{\rm wind} = v_z (^{12}{\rm CO})$. To calculate $\rho_{\rm mid}$, we use the gas surface density derived by \cite{zhang2021} (see Section \ref{sec:amb_driv_wind}) divided by $\sqrt{2\pi}H(r)$: $\rho_{\rm mid} (r)$ = $\Sigma(r)/{\sqrt{2 \pi} H(r)}$. With this, we estimate the mass loss rate via winds to be $4.4 \times 10^{-8}~M_\odot~{\rm yr}^{-1}$. In reality, the gas density of the wind can be smaller than the midplane density. In AS~209, the $^{12}$CO emission surface lies within $\approx$ 2 scale heights from the midplane, so assuming vertical hydrostatic equilibrium, the mass loss rate can be reduced by a factor of $e^{-2} \simeq 0.14$. Taking this into account, the mass loss rate via winds is $6.2 \times 10^{-9}~M_\odot~{\rm yr}^{-1}$. Calculating the total mass within the gap using the surface density from \cite{zhang2021}, we find that the gap would be depleted in a minimum $1.4\times10^4$ years if the mass loss rate is maintained and there is no gas radially advected into the gap.

Next, we estimate the mass inflow rate to the gap assuming an absence of winds using 
\begin{equation}
    \dot{M}_{\rm in} = 2 \pi r v_r \Sigma_{\rm in},
\end{equation}
where $v_r$ is the radial velocity of the inflowing gas and $\Sigma_{\rm in}$ is the surface density of the gas that falls into the gap. For a steady state viscous disk, the radial velocity $v_r$ can be described by $v_r = \alpha H \Omega_K$, where $\alpha$ is the coefficient characterizing the efficiency of the accretion (regardless of the origin), $H$ is the scale height, and $\Omega_K$ is the Keplerian orbital frequency. Adopting $H=14.3$~au and $\Sigma_{\rm in} = \Sigma_{\rm gas} \simeq 0.006~{\rm g~cm}^{-2}$ at 200~au (Figure 16 of \citealt{zhang2021}), we estimate a mass inflow rate of $3.0 \times 10^{-11} \times (\alpha/10^{-3})~M_\odot~{\rm yr}^{-1}$ which is smaller than $\dot{M}_{\rm wind}$ unless $\alpha \gtrsim 0.2$. This means that the strong upward flows seen in AS 209 can lead to mass loss from the gap, possibly halting the growth of the embedded planet and depleting the gas inside the gap.

Up to this point, our discussion has been focused on AS~209, but we can use the theory discussed to make a general scaling relation for wind-regulated terminal mass of giant planets. 
The depth of a gap opened by a planet can be described by
\begin{equation}
\label{eqn:sigma_gap}
    {\Sigma_{\rm gap} \over \Sigma} = {1 \over {1+0.04K}},
\end{equation}
where $\Sigma_{\rm gap}$ is the surface density at the center of the gap, $\Sigma$ is the unperturbed surface density, $K \equiv (M_p/M_*)^{2} (H/R)^{-5} \alpha^{-1}$ \citep{kanagawa2015}. For $K \gg 1$, applicable for planets opening a deep gap, we can write Equation (\ref{eqn:sigma_gap}) as
\begin{equation}
\label{eqn:sigma_gap2}
    {\Sigma_{\rm gap} \over \Sigma} \approx {1 \over 0.04K} = 0.25 \left({{M_p/M_*} \over {10^{-3}}}\right)^{-2} \left( {{H/R} \over {0.1}} \right)^5 \left( {\alpha \over 10^{-3}} \right).
\end{equation}

We can then relate the surface density at the gap center with the ambipolar Elsasser number as 
\begin{equation}
\label{eqn:am0}
    {\rm Am} = {{\gamma \rho_i} \over \Omega_K} = {\gamma \chi_i \rho \over \Omega_K} = {\gamma \chi_i \over \Omega_K} {\Sigma_{\rm gap} \over \sqrt{2\pi} H},
\end{equation}
where $\chi_i \equiv \rho_i/\rho$ is the ionization fraction.
Inserting $\Sigma_{\rm gap}$ from Equation (\ref{eqn:sigma_gap2}) into Equation (\ref{eqn:am0}) and re-organizing the equation, we obtain the terminal mass of a giant planet as follows:
\begin{eqnarray}
\label{eqn:terminal_mass}
    M_p\over M_*  & = & 9.7\times10^{-4} \left( {\rm Am} \over 0.1 \right)^{-1/2} \left( \chi_i \over 10^{-9} \right)^{1/2} \left( H/R \over 0.1 \right)^4 \nonumber \\
    & \times & \left( \alpha \over 10^{-3} \right)^{1/2}  \left( \Sigma \over 1~{\rm g~cm}^{-2} \right)^{1/2} \left( R \over 100~{\rm au} \right)^{1/4} \nonumber \\
    & \times & \left( M_* \over 1~M_\odot \right)^{-1/4} .
\end{eqnarray}
Using the fiducial parameters used in Equation (\ref{eqn:terminal_mass}) the terminal mass of a giant planet around a solar-mass star is about a Jupiter mass at 100~au.

Despite the prevalence of substructures in protoplanetary disks, attempts to search for young, forming planets through direct imaging resulted in a low detection rate \citep[see review by][and references therein]{benisty2022}. The properties of observed substructures  suggest that the majority of the young planet population has (sub-)Jovian mass \citep{bae2018,bae2022_ppvii, zhang_2018ApJ...869L..47Z,2019MNRAS.486..453L}. The wind-regulated terminal mass we estimated above coincides with the planet mass inferred from substructure properties, potentially helping to explain the dearth of directly imaged super-Jovian-mass young planets. Future kinematic studies of a larger sample of protoplanetary disks will enable us to test if wind-regulated growth of young planets is ubiquitous.

In the discussion above, we simplified the picture by assuming that there is no mass being fed to the CPD in the presence of large-scale winds. In hydrodynamic simulations, it is shown that the circumstellar disk gas can be supplied to the CPD through non-axisymmetric flows \citep{lubow1999}. Whether the same can happen in the presence of large-scale magnetically driven winds needs to be tested in the future, using non-ideal magnetohydrodynamic simulations with an embedded planet. If mass can still be supplied to the CPD in the presence of large-scale magnetically driven winds, the wind-regulated terminal mass in Equation (\ref{eqn:terminal_mass}) would provide a lower limit to the final mass of the planet.

\section{Summary} \label{sec:summary}

We have used $^{12}$CO, $^{13}$CO, and C$^{18}$O $J=2-1$ emission to carry out the detailed analysis of the kinematics within the AS 209 disk. We found significant perturbations in the rotational velocity in $^{12}$CO, up to $5\%$ of the Keplerian rotation, which is consistent with previous findings by \citet{as209_teague_2018ApJ...868..113T}. In addition to the perturbations in the rotational velocity, we found a strong {\it meridional fountain} (coherent upward flows) in $^{12}$CO and $^{13}$CO at $\simeq 1\farcs7$ (200~au). The upward flows are as fast as $150~{\rm m~s}^{-1}$ in $^{12}$CO, corresponding to about $50\%$ of the local sound speed or $6\%$ of the local Keplerian speed. Interestingly, these upward flows are co-located with an annular gap within which a candidate CPD is recently reported \citep{bae2022}. 

The observed upward flows are in the opposite direction to collapsing, downward flows within planet-carved gaps seen in hydrodynamic planet-disk interaction simulations, and are difficult to explain with an embedded planet alone. Instead, we propose a scenario in which the low density within the planet-carved gap has triggered magnetically driven winds via ambipolar diffusion. To support this idea, we estimated the ambipolar Elsasser number using the HCO$^+$ column density. At the radial location of the upward flows, we found that the ambipolar Elsasser number is about 0.1, broadly consistent with the value at which ambipolar diffusion drives strong winds in numerical simulations. In this scenario, we hypothesize that magnetically-driven winds from a planet-carved gap can limit/cease the growth of the planet embedded in the gap. This may be the explanation for the dearth of detections of gas-giant planets in disks with observed dust substructure with ALMA. We also provided a scaling relationship that describes the wind-regulated terminal mass. Using parameters generally applicable to protoplanetary disks, we found that the wind-regulated terminal mass around a solar-mass star is about a Jupiter mass at 100~au, which can explain the dearth of directly imaged super-Jovian-mass young planets at large orbital distances.

These results show compelling kinematic evidence of disk winds arising from the gap opened by a forming planet. 
In the future, constraining the ion density beyond HCO$^{+}$ will help better constrain the environment under which ambipolar diffusion-driven winds are launched. Observations constraining the morphology and strength of the magnetic fields in the AS~209 disk would help better understand the complex interplay between a forming planet and disk winds. Observations of species that can probe the warm outflowing gas from the low-density, higher regions, such as CI \citep{gressel2020,CIobs_2022ApJ...941L..24A}, could help further characterize the nature of the winds in the AS~209 disk. Kinematic studies for a larger sample of protoplanetary disks will help assess whether winds launched from planet-carved gaps are common or if the AS~209 disk is a unique case. Additionally, non-ideal magnetohydrodynamic planet-disk interaction simulations can prove (or dispute) if the activation of magnetically-driven winds within planet-carved gaps can regulate the growth of embedded planets. Finally, numerical studies of orbital migration in a disk with active winds will allow us to infer whether the CPD hosting planet in the AS~209 disk had formed at the current radial location or had formed at a different radial location but experienced inward/outward migration.

\begin{acknowledgments}
We thank Xiao Hu, Geoffroy Lesur, and  Gaylor Wafflard-Fernandez for helpful discussion. This paper makes use of the following ALMA data: \dataset[ADS/JAO.ALMA\#2018.1.01055.L]{https://almascience.nrao.edu/aq/?project\_code=2018.1.01055.L}. ALMA is a partnership of ESO (representing its member states), NSF (USA) and NINS (Japan), together with NRC (Canada), MOST and ASIAA (Taiwan), and KASI (Republic of Korea), in cooperation with the Republic of Chile. The Joint ALMA Observatory is operated by ESO, AUI/NRAO and NAOJ. The National Radio Astronomy Observatory is a facility of the National Science Foundation operated under cooperative agreement by Associated Universities, Inc. V.V.G. gratefully acknowledges support from FONDECYT Regular 1221352, ANID BASAL projects ACE210002 and FB210003, and ANID, -- Millennium Science Initiative Program -- NCN19\_171. Support for J. H. was provided by NASA through the NASA Hubble Fellowship grant \#HST-HF2-51460.001-A awarded by the Space Telescope Science Institute, which is operated by the Association of Universities for Research in Astronomy, Inc., for NASA, under contract NAS5-26555. R.L.G. acknowledges support from the Programme National “Physique et Chimie du Milieu Interstellaire” (PCMI) of CNRS/INSU with INC/INP co-funded by CEA and CNES. This project has received funding from the European Research Council (ERC) under the European Union’s Horizon 2020 research and innovation programme (PROTOPLANETS, grant agreement No. 101002188). C.W.~acknowledges financial support from the University of Leeds, the Science and Technology Facilities Council, and UK Research and Innovation (grant numbers ST/T000287/1 and MR/T040726/1).
This work was supported by a grant from the Simons Foundation (686302, KI\"O), and  an NSF AAG Grant (\#1907653, KI\"O). 
Support for F.L. was provided by NASA through the NASA Hubble Fellowship grant \#HST-HF2-51512.001-A awarded by the Space Telescope Science Institute, which is operated by the Association of Universities for Research in Astronomy, Incorporated, under NASA contract NAS5-26555. Y.A. acknowledges support by NAOJ ALMA Scientific Research Grant code 2019-13B. 
\end{acknowledgments}


\vspace{5mm}
\facilities{ALMA}

\software{\texttt{bettermoments} \citep{bettermoments}, \texttt{CASA} \citep{CASA}, \texttt{disksurf} \citep{disksurf}, \texttt{eddy} \citep{eddy}, Matplotlib \citep{matplotlib}, NumPy \citep{numpy}, SciPy \citep{scipy}}


\appendix 
\label{appendix}

\section{Results with Additional Emission Surface Models}
\label{sec:app_models}

\begin{figure}[h]
    \centering
    \includegraphics[width=1\textwidth]{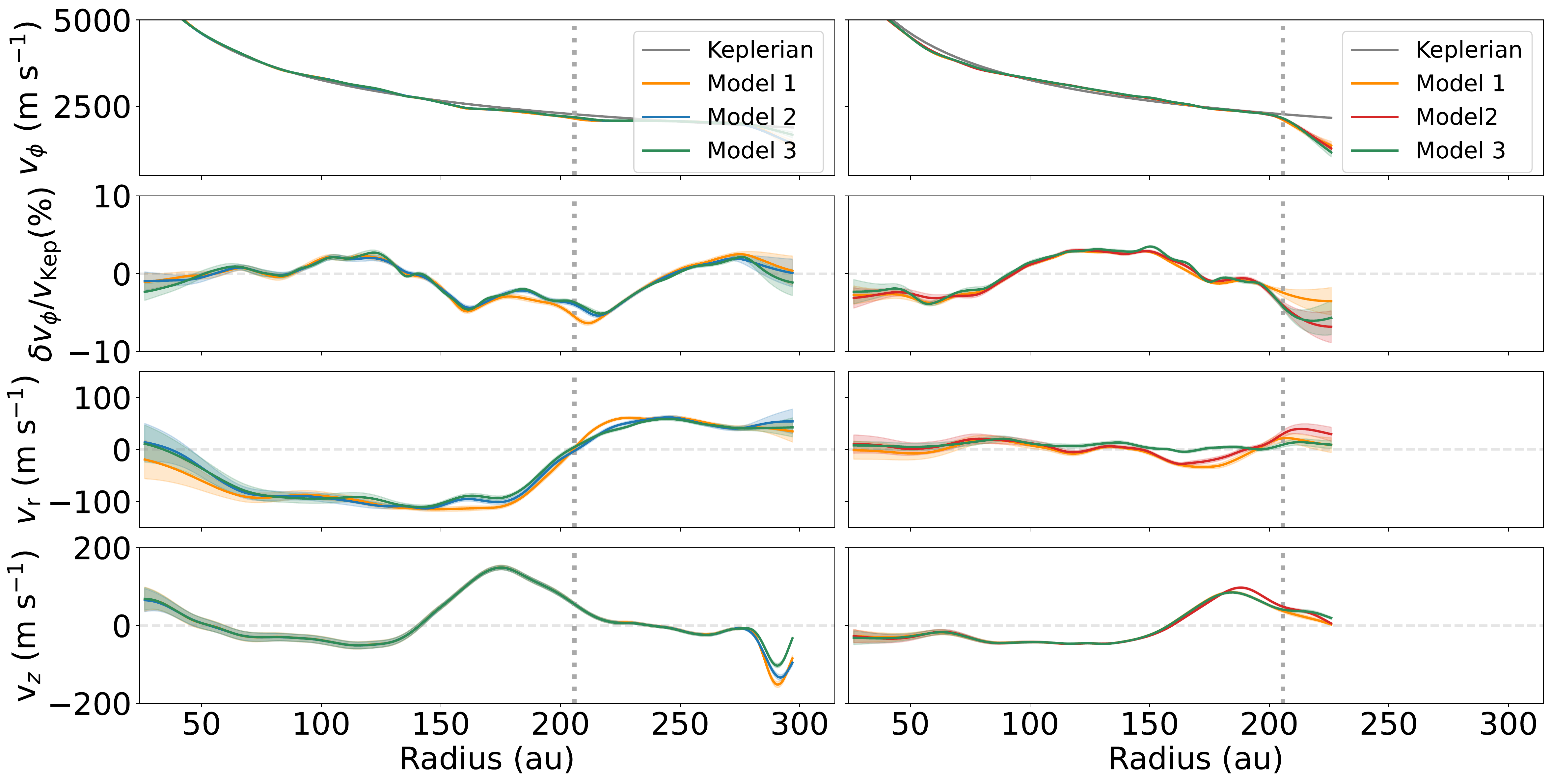}
    \caption{Comparison of velocity profiles found for (left) $^{12}$CO and (right) $^{13}$CO adopting different emission surface models. Orange: emission surfaces without a Gaussian dip (Model 1). Blue ($^{12}$CO) and red ($^{13}$CO): a tapered power law with a Gaussian gap (Model 2).  Green: Emission surface with Gaussian dip going to the midplane (Model 3). Note that the derived velocity profiles are insensitive to the emission surface models we adopt. The vertical dark grey dotted line shows the radial location of the CPD ($1\farcs7 \simeq 200$~au).
    }
    \label{fig:velocity_profiles_compare}
\end{figure}

In Tables \ref{tab:geometric_model1} and \ref{tab:geometric_model3}, we list $x_0$, $y_0$, PA, $M_*$, and $v_{\rm LSR}$ fitted with Model 1 and 3, respectively. Figure \ref{fig:velocity_profiles_compare} compares $^{12}$CO and $^{13}$CO velocity profiles for Models 1, 2, and 3. Note that the derived velocity profiles are insensitive to the emission surface models we adopt.

\begin{table*}[h]
\centering
\caption{Geometric properties calculated assuming a tapered power law for the emission surfaces (Model 1).}
\label{tab:geometric_model1}

\begin{tabular}{c|c|c|c|c|c}
         & x$_0$                   & y$_0$                  & PA                      & M$_*$                    & v$_{LSR}$                \\
\hline
          & (au)                    & (au)                   & ($^{\circ}$)              & (M$_{\odot}$)              & (km s$^{-1})$              \\
\hline
$^{12}$CO & -3.86$^{+0.08}_{-0.08}$ & 1.66$^{+0.12}_{-0.12}$ & 84.88$^{+0.07}_{-0.07}$ & 1.24$^{+0.003}_{-0.003}$ & 4.64$^{+0.001}_{-0.001}$ \\
$^{13}$CO & -3.82$^{+0.35}_{-0.36}$ & 0.71$^{+0.29}_{-0.29}$ & 86.03$^{+0.21}_{-0.21}$ & 1.25$^{+0.008}_{-0.007}$ & 4.65$^{+0.004}_{-0.004}$ \\
                 
\end{tabular}
\end{table*}

\begin{table*}[h]
\centering
\caption{Geometric properties calculated assuming a tapered power law with a Gaussian gap that reaches down to the midplane (Model 3).}
\label{tab:geometric_model3}

\begin{tabular}{c|c|c|c|c|c}
                   & x$_0$                   & y$_0$                  & PA                      & M$_*$                    & v$_{LSR}$                \\
\hline
          & (au)                    & (au)                   & ($^{\circ}$)              & (M$_{\odot}$)              & (km s$^{-1})$              \\
\hline
$^{12}$CO & -3.87$^{+0.08}_{-0.08}$ & 1.67$^{+0.12}_{-0.12}$ & 84.88$^{+0.07}_{-0.06}$ & 1.24$^{+0.003}_{-0.003}$ & 4.64$^{+0.001}_{-0.002}$  \\
$^{13}$CO & -1.67$^{+0.36}_{-0.36}$ & 0.73$^{+0.28}_{-0.28}$ & 86.01$^{+0.21}_{-0.21}$ & 1.25$^{+0.007}_{-0.008}$ & 4.65$^{+0.004}_{-0.004}$ \\

\end{tabular}
\end{table*}

\section{Results with JvM-uncorrected Cubes}
\label{sec:no_jvm}
\begin{figure*}
    \centering
    \includegraphics[width=\textwidth]{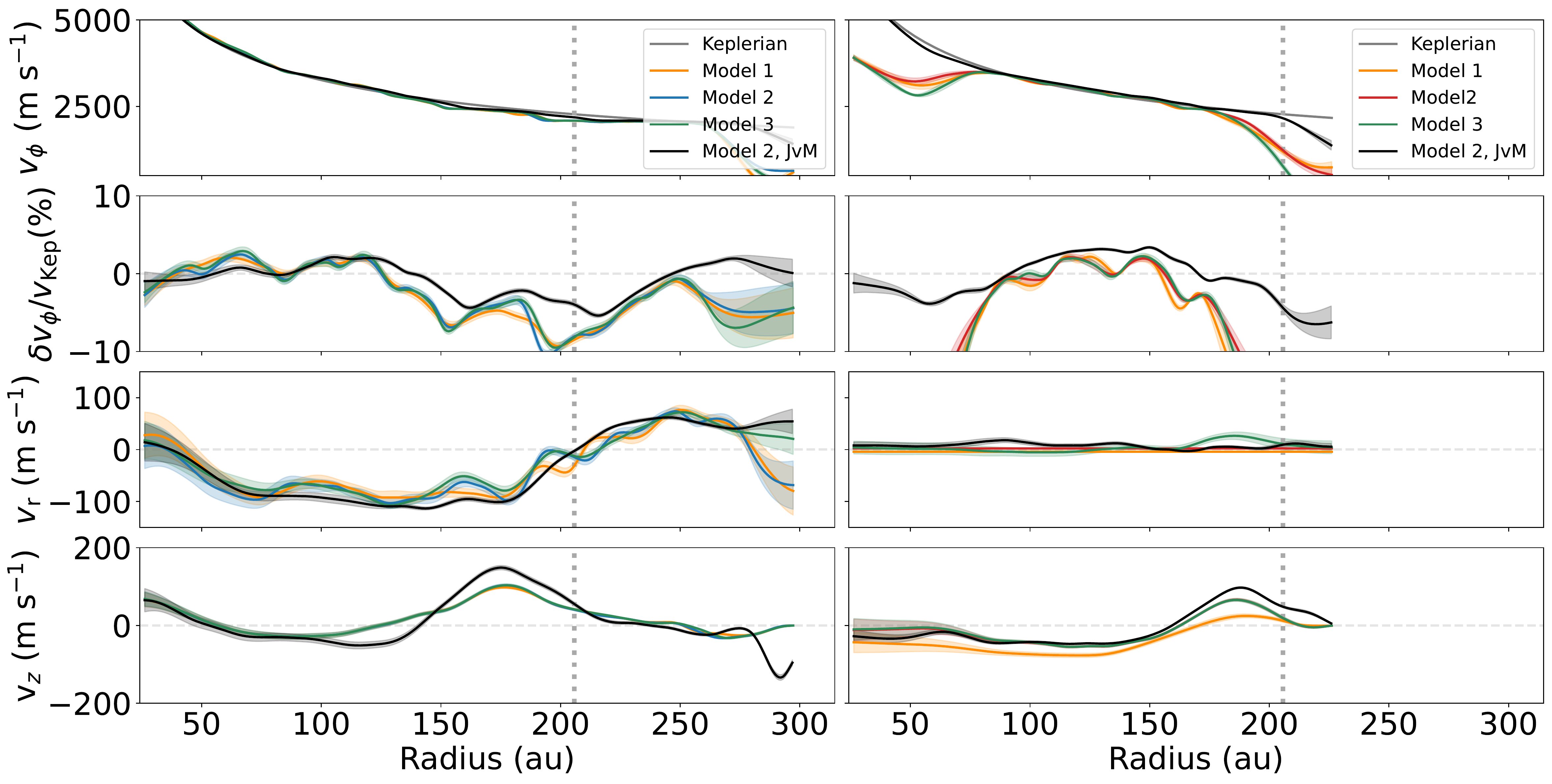}
    \caption{Same as Figure \ref{fig:vphi_vrad}, but with cubes that have not undergone JvM correction. Models 1, 2, and 3 represent the three emission surface Models shown in Figure \ref{fig:taperedemissionsurface}. The black curves are the velocity profiles for the JvM corrected cubes for comparison. We do not present C$^{18}$O because the signal-to-noise ratio if the JvM-uncorrected data is not sufficient for this velocity analysis. }
    \label{fig:vphi_vrad_noJVM}
\end{figure*}

In this section we repeat the velocity analysis presented in Section \ref{sec:analysis} but with JvM-uncorrected data cubes. For consistency, we use the emission surfaces and geometric properties derived from the JvM-corrected cubes. The resulting velocity profiles are shown in Figure \ref{fig:vphi_vrad_noJVM}. As shown in the figure, the inferred velocity profiles with the JvM-uncorrected data are broadly consistent with what we obtained with the JvM corrected data. Most importantly, the upward motions at $\sim150-200$~au are recovered.

\bibliography{sample631}{}

\begin{thebibliography}{}
\expandafter\ifx\csname natexlab\endcsname\relax\def\natexlab#1{#1}\fi
\providecommand{\url}[1]{\href{#1}{#1}}
\providecommand{\dodoi}[1]{doi:~\href{http://doi.org/#1}{\nolinkurl{#1}}}
\providecommand{\doeprint}[1]{\href{http://ascl.net/#1}{\nolinkurl{http://ascl.net/#1}}}
\providecommand{\doarXiv}[1]{\href{https://arxiv.org/abs/#1}{\nolinkurl{https://arxiv.org/abs/#1}}}

\bibitem[{{Aikawa} {et~al.}(2015){Aikawa}, {Furuya}, {Nomura}, \&
  {Qi}}]{aikawa2015}
{Aikawa}, Y., {Furuya}, K., {Nomura}, H., \& {Qi}, C. 2015, \apj, 807, 120,
  \dodoi{10.1088/0004-637X/807/2/120}

\bibitem[{{Aikawa} {et~al.}(2021){Aikawa}, {Cataldi}, {Yamato}, {Zhang},
  {Booth}, {Furuya}, {Andrews}, {Bae}, {Bergin}, {Bergner}, {Bosman},
  {Cleeves}, {Czekala}, {Guzm{\'a}n}, {Huang}, {Ilee}, {Law}, {Le Gal},
  {Loomis}, {M{\'e}nard}, {Nomura}, {{\"O}berg}, {Qi}, {Schwarz}, {Teague},
  {Tsukagoshi}, {Walsh}, \& {Wilner}}]{2021_Aikawa_ApJS..257...13A}
{Aikawa}, Y., {Cataldi}, G., {Yamato}, Y., {et~al.} 2021, \apjs, 257, 13,
  \dodoi{10.3847/1538-4365/ac143c}

\bibitem[{{Alarc{\'o}n} {et~al.}(2022){Alarc{\'o}n}, {Bergin}, \&
  {Teague}}]{CIobs_2022ApJ...941L..24A}
{Alarc{\'o}n}, F., {Bergin}, E.~A., \& {Teague}, R. 2022, \apjl, 941, L24,
  \dodoi{10.3847/2041-8213/aca6e6}

\bibitem[{{Andrews}(2020)}]{andrews2020}
{Andrews}, S.~M. 2020, \araa, 58, 483,
  \dodoi{10.1146/annurev-astro-031220-010302}

\bibitem[{{Andrews} {et~al.}(2009){Andrews}, {Wilner}, {Hughes}, {Qi}, \&
  {Dullemond}}]{2009ApJ...700.1502A}
{Andrews}, S.~M., {Wilner}, D.~J., {Hughes}, A.~M., {Qi}, C., \& {Dullemond},
  C.~P. 2009, \apj, 700, 1502, \dodoi{10.1088/0004-637X/700/2/1502}

\bibitem[{{Andrews} {et~al.}(2018){Andrews}, {Huang}, {P{\'e}rez}, {Isella},
  {Dullemond}, {Kurtovic}, {Guzm{\'a}n}, {Carpenter}, {Wilner}, {Zhang}, {Zhu},
  {Birnstiel}, {Bai}, {Benisty}, {Hughes}, {{\"O}berg}, \&
  {Ricci}}]{2018ApJ...869L..41A}
{Andrews}, S.~M., {Huang}, J., {P{\'e}rez}, L.~M., {et~al.} 2018, \apjl, 869,
  L41, \dodoi{10.3847/2041-8213/aaf741}

\bibitem[{{Bae} {et~al.}(2022{\natexlab{a}}){Bae}, {Isella}, {Zhu}, {Martin},
  {Okuzumi}, \& {Suriano}}]{bae2022_ppvii}
{Bae}, J., {Isella}, A., {Zhu}, Z., {et~al.} 2022{\natexlab{a}}, arXiv
  e-prints, arXiv:2210.13314.
\newblock \doarXiv{2210.13314}

\bibitem[{{Bae} {et~al.}(2018){Bae}, {Pinilla}, \& {Birnstiel}}]{bae2018}
{Bae}, J., {Pinilla}, P., \& {Birnstiel}, T. 2018, \apjl, 864, L26,
  \dodoi{10.3847/2041-8213/aadd51}

\bibitem[{{Bae} {et~al.}(2022{\natexlab{b}}){Bae}, {Teague}, {Andrews},
  {Benisty}, {Facchini}, {Galloway-Sprietsma}, {Loomis}, {Aikawa},
  {Alarc{\'o}n}, {Bergin}, {Bergner}, {Booth}, {Cataldi}, {Cleeves}, {Czekala},
  {Guzm{\'a}n}, {Huang}, {Ilee}, {Kurtovic}, {Law}, {Gal}, {Liu}, {Long},
  {M{\'e}nard}, {{\"O}berg}, {P{\'e}rez}, {Qi}, {Schwarz}, {Sierra}, {Walsh},
  {Wilner}, \& {Zhang}}]{bae2022}
{Bae}, J., {Teague}, R., {Andrews}, S.~M., {et~al.} 2022{\natexlab{b}}, \apjl,
  934, L20, \dodoi{10.3847/2041-8213/ac7fa3}

\bibitem[{{Bai}(2011)}]{bai2011}
{Bai}, X.-N. 2011, \apj, 739, 51, \dodoi{10.1088/0004-637X/739/1/51}

\bibitem[{{Bai} \& {Stone}(2011)}]{2011_bai_stone_ApJ...736..144B}
{Bai}, X.-N., \& {Stone}, J.~M. 2011, \apj, 736, 144,
  \dodoi{10.1088/0004-637X/736/2/144}

\bibitem[{{Bai} \& {Stone}(2013)}]{bai2013}
---. 2013, \apj, 769, 76, \dodoi{10.1088/0004-637X/769/1/76}

\bibitem[{{Barraza-Alfaro} {et~al.}(2021){Barraza-Alfaro}, {Flock}, {Marino},
  \& {P{\'e}rez}}]{barraza2021}
{Barraza-Alfaro}, M., {Flock}, M., {Marino}, S., \& {P{\'e}rez}, S. 2021, \aap,
  653, A113, \dodoi{10.1051/0004-6361/202140535}

\bibitem[{{Benisty} {et~al.}(2022){Benisty}, {Dominik}, {Follette}, {Garufi},
  {Ginski}, {Hashimoto}, {Keppler}, {Kley}, \& {Monnier}}]{benisty2022}
{Benisty}, M., {Dominik}, C., {Follette}, K., {et~al.} 2022, arXiv e-prints,
  arXiv:2203.09991.
\newblock \doarXiv{2203.09991}

\bibitem[{{B{\'e}thune} {et~al.}(2017){B{\'e}thune}, {Lesur}, \&
  {Ferreira}}]{bethune2017}
{B{\'e}thune}, W., {Lesur}, G., \& {Ferreira}, J. 2017, \aap, 600, A75,
  \dodoi{10.1051/0004-6361/201630056}

\bibitem[{{Cieza} {et~al.}(2021){Cieza}, {Gonz{\'a}lez-Ruilova}, {Hales},
  {Pinilla}, {Ru{\'\i}z-Rodr{\'\i}guez}, {Zurlo}, {Casassus}, {P{\'e}rez},
  {C{\'a}novas}, {Arce-Tord}, {Flock}, {Kurtovic}, {Marino}, {Nogueira},
  {Perez}, {Price}, {Principe}, \& {Williams}}]{2021MNRAS.501.2934C}
{Cieza}, L.~A., {Gonz{\'a}lez-Ruilova}, C., {Hales}, A.~S., {et~al.} 2021,
  \mnras, 501, 2934, \dodoi{10.1093/mnras/staa3787}

\bibitem[{{Currie} {et~al.}(2022){Currie}, {Lawson}, {Schneider}, {Lyra},
  {Wisniewski}, {Grady}, {Guyon}, {Tamura}, {Kotani}, {Kawahara}, {Brandt},
  {Uyama}, {Muto}, {Dong}, {Kudo}, {Hashimoto}, {Fukagawa}, {Wagner}, {Lozi},
  {Chilcote}, {Tobin}, {Groff}, {Ward-Duong}, {Januszewski}, {Norris},
  {Tuthill}, {van der Marel}, {Sitko}, {Deo}, {Vievard}, {Jovanovic},
  {Martinache}, \& {Skaf}}]{2022NatAs...6..751C}
{Currie}, T., {Lawson}, K., {Schneider}, G., {et~al.} 2022, Nature Astronomy,
  6, 751, \dodoi{10.1038/s41550-022-01634-x}

\bibitem[{{Czekala} {et~al.}(2021){Czekala}, {Loomis}, {Teague}, {Booth},
  {Huang}, {Cataldi}, {Ilee}, {Law}, {Walsh}, {Bosman}, {Guzm{\'a}n}, {Gal},
  {{\"O}berg}, {Yamato}, {Aikawa}, {Andrews}, {Bae}, {Bergin}, {Bergner},
  {Cleeves}, {Kurtovic}, {M{\'e}nard}, {Nomura}, {P{\'e}rez}, {Qi}, {Schwarz},
  {Tsukagoshi}, {Waggoner}, {Wilner}, \& {Zhang}}]{czekala_2021ApJS..257....2C}
{Czekala}, I., {Loomis}, R.~A., {Teague}, R., {et~al.} 2021, \apjs, 257, 2,
  \dodoi{10.3847/1538-4365/ac1430}

\bibitem[{{Draine}(2011)}]{draine2011}
{Draine}, B.~T. 2011, {Physics of the Interstellar and Intergalactic Medium}

\bibitem[{{Fedele} {et~al.}(2018){Fedele}, {Tazzari}, {Booth}, {Testi},
  {Clarke}, {Pascucci}, {Kospal}, {Semenov}, {Bruderer}, {Henning}, \&
  {Teague}}]{fedele_2018A&A...610A..24F}
{Fedele}, D., {Tazzari}, M., {Booth}, R., {et~al.} 2018, \aap, 610, A24,
  \dodoi{10.1051/0004-6361/201731978}

\bibitem[{{Foreman-Mackey} {et~al.}(2017){Foreman-Mackey}, {Agol},
  {Ambikasaran}, \& {Angus}}]{2017AJ....154..220F}
{Foreman-Mackey}, D., {Agol}, E., {Ambikasaran}, S., \& {Angus}, R. 2017, \aj,
  154, 220, \dodoi{10.3847/1538-3881/aa9332}

\bibitem[{{Foreman-Mackey} {et~al.}(2013){Foreman-Mackey}, {Hogg}, {Lang}, \&
  {Goodman}}]{2013PASP..125..306F}
{Foreman-Mackey}, D., {Hogg}, D.~W., {Lang}, D., \& {Goodman}, J. 2013, \pasp,
  125, 306, \dodoi{10.1086/670067}

\bibitem[{{Fung} \& {Chiang}(2016)}]{fung_2016ApJ...832..105F}
{Fung}, J., \& {Chiang}, E. 2016, \apj, 832, 105,
  \dodoi{10.3847/0004-637X/832/2/105}

\bibitem[{{Gaia Collaboration} {et~al.}(2021){Gaia Collaboration}, {Brown},
  {Vallenari}, {Prusti}, {de Bruijne}, {Babusiaux}, {Biermann}, {Creevey},
  {Evans}, {Eyer}, {Hutton}, {Jansen}, {Jordi}, {Klioner}, {Lammers},
  {Lindegren}, {Luri}, {Mignard}, {Panem}, {Pourbaix}, {Randich}, {Sartoretti},
  {Soubiran}, {Walton}, {Arenou}, {Bailer-Jones}, {Bastian}, {Cropper},
  {Drimmel}, {Katz}, {Lattanzi}, {van Leeuwen}, {Bakker}, {Cacciari},
  {Casta{\~n}eda}, {De Angeli}, {Ducourant}, {Fabricius}, {Fouesneau},
  {Fr{\'e}mat}, {Guerra}, {Guerrier}, {Guiraud}, {Jean-Antoine Piccolo},
  {Masana}, {Messineo}, {Mowlavi}, {Nicolas}, {Nienartowicz}, {Pailler},
  {Panuzzo}, {Riclet}, {Roux}, {Seabroke}, {Sordo}, {Tanga}, {Th{\'e}venin},
  {Gracia-Abril}, {Portell}, {Teyssier}, {Altmann}, {Andrae}, {Bellas-Velidis},
  {Benson}, {Berthier}, {Blomme}, {Brugaletta}, {Burgess}, {Busso}, {Carry},
  {Cellino}, {Cheek}, {Clementini}, {Damerdji}, {Davidson}, {Delchambre},
  {Dell'Oro}, {Fern{\'a}ndez-Hern{\'a}ndez}, {Galluccio}, {Garc{\'\i}a-Lario},
  {Garcia-Reinaldos}, {Gonz{\'a}lez-N{\'u}{\~n}ez}, {Gosset}, {Haigron},
  {Halbwachs}, {Hambly}, {Harrison}, {Hatzidimitriou}, {Heiter},
  {Hern{\'a}ndez}, {Hestroffer}, {Hodgkin}, {Holl}, {Jan{\ss}en}, {Jevardat de
  Fombelle}, {Jordan}, {Krone-Martins}, {Lanzafame}, {L{\"o}ffler}, {Lorca},
  {Manteiga}, {Marchal}, {Marrese}, {Moitinho}, {Mora}, {Muinonen}, {Osborne},
  {Pancino}, {Pauwels}, {Petit}, {Recio-Blanco}, {Richards}, {Riello},
  {Rimoldini}, {Robin}, {Roegiers}, {Rybizki}, {Sarro}, {Siopis}, {Smith},
  {Sozzetti}, {Ulla}, {Utrilla}, {van Leeuwen}, {van Reeven}, {Abbas}, {Abreu
  Aramburu}, {Accart}, {Aerts}, {Aguado}, {Ajaj}, {Altavilla}, {{\'A}lvarez},
  {{\'A}lvarez Cid-Fuentes}, {Alves}, {Anderson}, {Anglada Varela}, {Antoja},
  {Audard}, {Baines}, {Baker}, {Balaguer-N{\'u}{\~n}ez}, {Balbinot}, {Balog},
  {Barache}, {Barbato}, {Barros}, {Barstow}, {Bartolom{\'e}}, {Bassilana},
  {Bauchet}, {Baudesson-Stella}, {Becciani}, {Bellazzini}, {Bernet}, {Bertone},
  {Bianchi}, {Blanco-Cuaresma}, {Boch}, {Bombrun}, {Bossini}, {Bouquillon},
  {Bragaglia}, {Bramante}, {Breedt}, {Bressan}, {Brouillet}, {Bucciarelli},
  {Burlacu}, {Busonero}, {Butkevich}, {Buzzi}, {Caffau}, {Cancelliere},
  {C{\'a}novas}, {Cantat-Gaudin}, {Carballo}, {Carlucci}, {Carnerero},
  {Carrasco}, {Casamiquela}, {Castellani}, {Castro-Ginard}, {Castro Sampol},
  {Chaoul}, {Charlot}, {Chemin}, {Chiavassa}, {Cioni}, {Comoretto}, {Cooper},
  {Cornez}, {Cowell}, {Crifo}, {Crosta}, {Crowley}, {Dafonte}, {Dapergolas},
  {David}, {David}, {de Laverny}, {De Luise}, {De March}, {De Ridder}, {de
  Souza}, {de Teodoro}, {de Torres}, {del Peloso}, {del Pozo}, {Delbo},
  {Delgado}, {Delgado}, {Delisle}, {Di Matteo}, {Diakite}, {Diener},
  {Distefano}, {Dolding}, {Eappachen}, {Edvardsson}, {Enke}, {Esquej}, {Fabre},
  {Fabrizio}, {Faigler}, {Fedorets}, {Fernique}, {Fienga}, {Figueras},
  {Fouron}, {Fragkoudi}, {Fraile}, {Franke}, {Gai}, {Garabato},
  {Garcia-Gutierrez}, {Garc{\'\i}a-Torres}, {Garofalo}, {Gavras}, {Gerlach},
  {Geyer}, {Giacobbe}, {Gilmore}, {Girona}, {Giuffrida}, {Gomel}, {Gomez},
  {Gonzalez-Santamaria}, {Gonz{\'a}lez-Vidal}, {Granvik},
  {Guti{\'e}rrez-S{\'a}nchez}, {Guy}, {Hauser}, {Haywood}, {Helmi}, {Hidalgo},
  {Hilger}, {H{\l}adczuk}, {Hobbs}, {Holland}, {Huckle}, {Jasniewicz},
  {Jonker}, {Juaristi Campillo}, {Julbe}, {Karbevska}, {Kervella}, {Khanna},
  {Kochoska}, {Kontizas}, {Kordopatis}, {Korn}, {Kostrzewa-Rutkowska},
  {Kruszy{\'n}ska}, {Lambert}, {Lanza}, {Lasne}, {Le Campion}, {Le Fustec},
  {Lebreton}, {Lebzelter}, {Leccia}, {Leclerc}, {Lecoeur-Taibi}, {Liao},
  {Licata}, {Lindstr{\o}m}, {Lister}, {Livanou}, {Lobel}, {Madrero Pardo},
  {Managau}, {Mann}, {Marchant}, {Marconi}, {Marcos Santos}, {Marinoni},
  {Marocco}, {Marshall}, {Martin Polo}, {Mart{\'\i}n-Fleitas}, {Masip},
  {Massari}, {Mastrobuono-Battisti}, {Mazeh}, {McMillan}, {Messina},
  {Michalik}, {Millar}, {Mints}, {Molina}, {Molinaro}, {Moln{\'a}r},
  {Montegriffo}, {Mor}, {Morbidelli}, {Morel}, {Morris}, {Mulone}, {Munoz},
  {Muraveva}, {Murphy}, {Musella}, {Noval}, {Ord{\'e}novic}, {Orr{\`u}},
  {Osinde}, {Pagani}, {Pagano}, {Palaversa}, {Palicio}, {Panahi}, {Pawlak},
  {Pe{\~n}alosa Esteller}, {Penttil{\"a}}, {Piersimoni}, {Pineau}, {Plachy},
  {Plum}, {Poggio}, {Poretti}, {Poujoulet}, {Pr{\v{s}}a}, {Pulone}, {Racero},
  {Ragaini}, {Rainer}, {Raiteri}, {Rambaux}, {Ramos}, {Ramos-Lerate}, {Re
  Fiorentin}, {Regibo}, {Reyl{\'e}}, {Ripepi}, {Riva}, {Rixon}, {Robichon},
  {Robin}, {Roelens}, {Rohrbasser}, {Romero-G{\'o}mez}, {Rowell}, {Royer},
  {Rybicki}, {Sadowski}, {Sagrist{\`a} Sell{\'e}s}, {Sahlmann}, {Salgado},
  {Salguero}, {Samaras}, {Sanchez Gimenez}, {Sanna}, {Santove{\~n}a},
  {Sarasso}, {Schultheis}, {Sciacca}, {Segol}, {Segovia}, {S{\'e}gransan},
  {Semeux}, {Shahaf}, {Siddiqui}, {Siebert}, {Siltala}, {Slezak}, {Smart},
  {Solano}, {Solitro}, {Souami}, {Souchay}, {Spagna}, {Spoto}, {Steele},
  {Steidelm{\"u}ller}, {Stephenson}, {S{\"u}veges}, {Szabados}, {Szegedi-Elek},
  {Taris}, {Tauran}, {Taylor}, {Teixeira}, {Thuillot}, {Tonello}, {Torra},
  {Torra}, {Turon}, {Unger}, {Vaillant}, {van Dillen}, {Vanel}, {Vecchiato},
  {Viala}, {Vicente}, {Voutsinas}, {Weiler}, {Wevers}, {Wyrzykowski}, {Yoldas},
  {Yvard}, {Zhao}, {Zorec}, {Zucker}, {Zurbach}, \& {Zwitter}}]{gaia_2021}
{Gaia Collaboration}, {Brown}, A.~G.~A., {Vallenari}, A., {et~al.} 2021, \aap,
  650, C3, \dodoi{10.1051/0004-6361/202039657e}

\bibitem[{{Gressel} {et~al.}(2013){Gressel}, {Nelson}, {Turner}, \&
  {Ziegler}}]{gressel2013}
{Gressel}, O., {Nelson}, R.~P., {Turner}, N.~J., \& {Ziegler}, U. 2013, \apj,
  779, 59, \dodoi{10.1088/0004-637X/779/1/59}

\bibitem[{{Gressel} {et~al.}(2020){Gressel}, {Ramsey}, {Brinch}, {Nelson},
  {Turner}, \& {Bruderer}}]{gressel2020}
{Gressel}, O., {Ramsey}, J.~P., {Brinch}, C., {et~al.} 2020, \apj, 896, 126,
  \dodoi{10.3847/1538-4357/ab91b7}

\bibitem[{{Gressel} {et~al.}(2015){Gressel}, {Turner}, {Nelson}, \&
  {McNally}}]{gressel2015}
{Gressel}, O., {Turner}, N.~J., {Nelson}, R.~P., \& {McNally}, C.~P. 2015,
  \apj, 801, 84, \dodoi{10.1088/0004-637X/801/2/84}

\bibitem[{{Guzm{\'a}n} {et~al.}(2018){Guzm{\'a}n}, {Huang}, {Andrews},
  {Isella}, {P{\'e}rez}, {Carpenter}, {Dullemond}, {Ricci}, {Birnstiel},
  {Zhang}, {Zhu}, {Bai}, {Benisty}, {{\"O}berg}, \&
  {Wilner}}]{guzman_2018ApJ...869L..48G}
{Guzm{\'a}n}, V.~V., {Huang}, J., {Andrews}, S.~M., {et~al.} 2018, \apjl, 869,
  L48, \dodoi{10.3847/2041-8213/aaedae}

\bibitem[{{Haffert} {et~al.}(2019){Haffert}, {Bohn}, {de Boer}, {Snellen},
  {Brinchmann}, {Girard}, {Keller}, \& {Bacon}}]{haffert2019}
{Haffert}, S.~Y., {Bohn}, A.~J., {de Boer}, J., {et~al.} 2019, Nature
  Astronomy, 3, 749, \dodoi{10.1038/s41550-019-0780-5}

\bibitem[{{Harrison} {et~al.}(2021){Harrison}, {Looney}, {Stephens}, {Li},
  {Teague}, {Crutcher}, {Yang}, {Cox}, {Fern{\'a}ndez-L{\'o}pez}, \&
  {Shinnaga}}]{2021_zeeman_mag_ApJ...908..141H}
{Harrison}, R.~E., {Looney}, L.~W., {Stephens}, I.~W., {et~al.} 2021, \apj,
  908, 141, \dodoi{10.3847/1538-4357/abd94e}

\bibitem[{{Hu} {et~al.}(2022){Hu}, {Li}, {Zhu}, \&
  {Yang}}]{2022_hu_MNRAS.516.2006H}
{Hu}, X., {Li}, Z.-Y., {Zhu}, Z., \& {Yang}, C.-C. 2022, \mnras, 516, 2006,
  \dodoi{10.1093/mnras/stac1799}

\bibitem[{{Huang} {et~al.}(2016){Huang}, {{\"O}berg}, \&
  {Andrews}}]{huang_2016ApJ...823L..18H}
{Huang}, J., {{\"O}berg}, K.~I., \& {Andrews}, S.~M. 2016, \apjl, 823, L18,
  \dodoi{10.3847/2041-8205/823/1/L18}

\bibitem[{{Huang} {et~al.}(2018){Huang}, {Andrews}, {Dullemond}, {Isella},
  {P{\'e}rez}, {Guzm{\'a}n}, {{\"O}berg}, {Zhu}, {Zhang}, {Bai}, {Benisty},
  {Birnstiel}, {Carpenter}, {Hughes}, {Ricci}, {Weaver}, \&
  {Wilner}}]{huang_2018ApJ...869L..42H}
{Huang}, J., {Andrews}, S.~M., {Dullemond}, C.~P., {et~al.} 2018, \apjl, 869,
  L42, \dodoi{10.3847/2041-8213/aaf740}

\bibitem[{{Hunter}(2007)}]{matplotlib}
{Hunter}, J.~D. 2007, Computing in Science and Engineering, 9, 90,
  \dodoi{10.1109/MCSE.2007.55}

\bibitem[{{Izquierdo} {et~al.}(2023){Izquierdo}, {Testi}, {Facchini},
  {Rosotti}, {van Dishoeck}, {W{\"o}lfer}, \&
  {Paneque-Carre{\~n}o}}]{Izquierdo2023}
{Izquierdo}, A., {Testi}, L., {Facchini}, S., {et~al.} 2023, \aap, accepted

\bibitem[{{Izquierdo} {et~al.}(2021){Izquierdo}, {Testi}, {Facchini},
  {Rosotti}, \& {van Dishoeck}}]{izquierdo2021}
{Izquierdo}, A.~F., {Testi}, L., {Facchini}, S., {Rosotti}, G.~P., \& {van
  Dishoeck}, E.~F. 2021, \aap, 650, A179, \dodoi{10.1051/0004-6361/202140779}

\bibitem[{{Jorsater} \& {van Moorsel}(1995)}]{1995AJ....110.2037J}
{Jorsater}, S., \& {van Moorsel}, G.~A. 1995, \aj, 110, 2037,
  \dodoi{10.1086/117668}

\bibitem[{{Kanagawa} {et~al.}(2015){Kanagawa}, {Muto}, {Tanaka}, {Tanigawa},
  {Takeuchi}, {Tsukagoshi}, \& {Momose}}]{kanagawa2015}
{Kanagawa}, K.~D., {Muto}, T., {Tanaka}, H., {et~al.} 2015, \apjl, 806, L15,
  \dodoi{10.1088/2041-8205/806/1/L15}

\bibitem[{{Keppler} {et~al.}(2018){Keppler}, {Benisty}, {M{\"u}ller},
  {Henning}, {van Boekel}, {Cantalloube}, {Ginski}, {van Holstein}, {Maire},
  {Pohl}, {Samland}, {Avenhaus}, {Baudino}, {Boccaletti}, {de Boer},
  {Bonnefoy}, {Chauvin}, {Desidera}, {Langlois}, {Lazzoni}, {Marleau},
  {Mordasini}, {Pawellek}, {Stolker}, {Vigan}, {Zurlo}, {Birnstiel},
  {Brandner}, {Feldt}, {Flock}, {Girard}, {Gratton}, {Hagelberg}, {Isella},
  {Janson}, {Juhasz}, {Kemmer}, {Kral}, {Lagrange}, {Launhardt}, {Matter},
  {M{\'e}nard}, {Milli}, {Molli{\`e}re}, {Olofsson}, {P{\'e}rez}, {Pinilla},
  {Pinte}, {Quanz}, {Schmidt}, {Udry}, {Wahhaj}, {Williams}, {Buenzli},
  {Cudel}, {Dominik}, {Galicher}, {Kasper}, {Lannier}, {Mesa}, {Mouillet},
  {Peretti}, {Perrot}, {Salter}, {Sissa}, {Wildi}, {Abe}, {Antichi},
  {Augereau}, {Baruffolo}, {Baudoz}, {Bazzon}, {Beuzit}, {Blanchard}, {Brems},
  {Buey}, {De Caprio}, {Carbillet}, {Carle}, {Cascone}, {Cheetham}, {Claudi},
  {Costille}, {Delboulb{\'e}}, {Dohlen}, {Fantinel}, {Feautrier}, {Fusco},
  {Giro}, {Gluck}, {Gry}, {Hubin}, {Hugot}, {Jaquet}, {Le Mignant}, {Llored},
  {Madec}, {Magnard}, {Martinez}, {Maurel}, {Meyer}, {M{\"o}ller-Nilsson},
  {Moulin}, {Mugnier}, {Orign{\'e}}, {Pavlov}, {Perret}, {Petit}, {Pragt},
  {Puget}, {Rabou}, {Ramos}, {Rigal}, {Rochat}, {Roelfsema}, {Rousset}, {Roux},
  {Salasnich}, {Sauvage}, {Sevin}, {Soenke}, {Stadler}, {Suarez}, {Turatto}, \&
  {Weber}}]{keppler2018}
{Keppler}, M., {Benisty}, M., {M{\"u}ller}, A., {et~al.} 2018, \aap, 617, A44,
  \dodoi{10.1051/0004-6361/201832957}

\bibitem[{{Kley}(1999)}]{kley1999}
{Kley}, W. 1999, \mnras, 303, 696, \dodoi{10.1046/j.1365-8711.1999.02198.x}

\bibitem[{{Kley} {et~al.}(2001){Kley}, {D'Angelo}, \& {Henning}}]{kley2001}
{Kley}, W., {D'Angelo}, G., \& {Henning}, T. 2001, \apj, 547, 457,
  \dodoi{10.1086/318345}

\bibitem[{{Law} {et~al.}(2021{\natexlab{a}}){Law}, {Loomis}, {Teague},
  {{\"O}berg}, {Czekala}, {Andrews}, {Huang}, {Aikawa}, {Alarc{\'o}n}, {Bae},
  {Bergin}, {Bergner}, {Boehler}, {Booth}, {Bosman}, {Calahan}, {Cataldi},
  {Cleeves}, {Furuya}, {Guzm{\'a}n}, {Ilee}, {Le Gal}, {Liu}, {Long},
  {M{\'e}nard}, {Nomura}, {Qi}, {Schwarz}, {Sierra}, {Tsukagoshi}, {Yamato},
  {van't Hoff}, {Walsh}, {Wilner}, \& {Zhang}}]{law2021a}
{Law}, C.~J., {Loomis}, R.~A., {Teague}, R., {et~al.} 2021{\natexlab{a}},
  \apjs, 257, 3, \dodoi{10.3847/1538-4365/ac1434}

\bibitem[{{Law} {et~al.}(2021{\natexlab{b}}){Law}, {Teague}, {Loomis}, {Bae},
  {{\"O}berg}, {Czekala}, {Andrews}, {Aikawa}, {Alarc{\'o}n}, {Bergin},
  {Bergner}, {Booth}, {Bosman}, {Calahan}, {Cataldi}, {Cleeves}, {Furuya},
  {Guzm{\'a}n}, {Huang}, {Ilee}, {Le Gal}, {Liu}, {Long}, {M{\'e}nard},
  {Nomura}, {P{\'e}rez}, {Qi}, {Schwarz}, {Soto}, {Tsukagoshi}, {Yamato},
  {van't Hoff}, {Walsh}, {Wilner}, \& {Zhang}}]{Law_2021}
{Law}, C.~J., {Teague}, R., {Loomis}, R.~A., {et~al.} 2021{\natexlab{b}},
  \apjs, 257, 4, \dodoi{10.3847/1538-4365/ac1439}

\bibitem[{{Law} {et~al.}(2022{\natexlab{a}}){Law}, {Crystian}, {Teague},
  {{\"O}berg}, {Rich}, {Andrews}, {Bae}, {Flaherty}, {Guzm{\'a}n}, {Huang},
  {Ilee}, {Kastner}, {Loomis}, {Long}, {P{\'e}rez}, {P{\'e}rez}, {Qi},
  {Rosotti}, {Ru{\'\i}z-Rodr{\'\i}guez}, {Tsukagoshi}, \& {Wilner}}]{Law_2022}
{Law}, C.~J., {Crystian}, S., {Teague}, R., {et~al.} 2022{\natexlab{a}}, \apj,
  932, 114, \dodoi{10.3847/1538-4357/ac6c02}

\bibitem[{{Law} {et~al.}(2022{\natexlab{b}}){Law}, {Teague}, {{\"O}berg},
  {Rich}, {Andrews}, {Bae}, {Benisty}, {Facchini}, {Flaherty}, {Isella}, {Jin},
  {Hashimoto}, {Huang}, {Loomis}, {Long}, {Mu{\~n}oz-Romero},
  {Paneque-Carre{\~n}o}, {P{\'e}rez}, {Qi}, {Schwarz}, {Stadler}, {Tsukagoshi},
  {Wilner}, \& {van der Plas}}]{2022arXiv221208667L}
{Law}, C.~J., {Teague}, R., {{\"O}berg}, K.~I., {et~al.} 2022{\natexlab{b}},
  arXiv e-prints, arXiv:2212.08667.
\newblock \doarXiv{2212.08667}

\bibitem[{{Lesur} {et~al.}(2022){Lesur}, {Ercolano}, {Flock}, {Lin}, {Yang},
  {Barranco}, {Benitez-Llambay}, {Goodman}, {Johansen}, {Klahr}, {Laibe},
  {Lyra}, {Marcus}, {Nelson}, {Squire}, {Simon}, {Turner}, {Umurhan}, \&
  {Youdin}}]{lesur2022}
{Lesur}, G., {Ercolano}, B., {Flock}, M., {et~al.} 2022, arXiv e-prints,
  arXiv:2203.09821.
\newblock \doarXiv{2203.09821}

\bibitem[{{Lodato} {et~al.}(2019){Lodato}, {Dipierro}, {Ragusa}, {Long},
  {Herczeg}, {Pascucci}, {Pinilla}, {Manara}, {Tazzari}, {Liu}, {Mulders},
  {Harsono}, {Boehler}, {M{\'e}nard}, {Johnstone}, {Salyk}, {van der Plas},
  {Cabrit}, {Edwards}, {Fischer}, {Hendler}, {Nisini}, {Rigliaco}, {Avenhaus},
  {Banzatti}, \& {Gully-Santiago}}]{2019MNRAS.486..453L}
{Lodato}, G., {Dipierro}, G., {Ragusa}, E., {et~al.} 2019, \mnras, 486, 453,
  \dodoi{10.1093/mnras/stz913}

\bibitem[{{Long} {et~al.}(2018){Long}, {Pinilla}, {Herczeg}, {Harsono},
  {Dipierro}, {Pascucci}, {Hendler}, {Tazzari}, {Ragusa}, {Salyk}, {Edwards},
  {Lodato}, {van de Plas}, {Johnstone}, {Liu}, {Boehler}, {Cabrit}, {Manara},
  {Menard}, {Mulders}, {Nisini}, {Fischer}, {Rigliaco}, {Banzatti}, {Avenhaus},
  \& {Gully-Santiago}}]{2018ApJ...869...17L}
{Long}, F., {Pinilla}, P., {Herczeg}, G.~J., {et~al.} 2018, \apj, 869, 17,
  \dodoi{10.3847/1538-4357/aae8e1}

\bibitem[{{Lubow} {et~al.}(1999){Lubow}, {Seibert}, \&
  {Artymowicz}}]{lubow1999}
{Lubow}, S.~H., {Seibert}, M., \& {Artymowicz}, P. 1999, \apj, 526, 1001,
  \dodoi{10.1086/308045}

\bibitem[{{McMullin} {et~al.}(2007){McMullin}, {Waters}, {Schiebel}, {Young},
  \& {Golap}}]{CASA}
{McMullin}, J.~P., {Waters}, B., {Schiebel}, D., {Young}, W., \& {Golap}, K.
  2007, in Astronomical Society of the Pacific Conference Series, Vol. 376,
  Astronomical Data Analysis Software and Systems XVI, ed. R.~A. {Shaw},
  F.~{Hill}, \& D.~J. {Bell}, 127

\bibitem[{{Morbidelli} {et~al.}(2014){Morbidelli}, {Szul{\'a}gyi}, {Crida},
  {Lega}, {Bitsch}, {Tanigawa}, \& {Kanagawa}}]{2014Icar..232..266M}
{Morbidelli}, A., {Szul{\'a}gyi}, J., {Crida}, A., {et~al.} 2014, \icarus, 232,
  266, \dodoi{10.1016/j.icarus.2014.01.010}

\bibitem[{{Nelson} {et~al.}(2013){Nelson}, {Gressel}, \&
  {Umurhan}}]{nelson2013}
{Nelson}, R.~P., {Gressel}, O., \& {Umurhan}, O.~M. 2013, \mnras, 435, 2610,
  \dodoi{10.1093/mnras/stt1475}

\bibitem[{{{\"O}berg} {et~al.}(2011){{\"O}berg}, {Qi}, {Fogel}, {Bergin},
  {Andrews}, {Espaillat}, {Wilner}, {Pascucci}, \&
  {Kastner}}]{oberg_2011ApJ...734...98O}
{{\"O}berg}, K.~I., {Qi}, C., {Fogel}, J. K.~J., {et~al.} 2011, \apj, 734, 98,
  \dodoi{10.1088/0004-637X/734/2/98}

\bibitem[{{{\"O}berg} {et~al.}(2021){{\"O}berg}, {Guzm{\'a}n}, {Walsh},
  {Aikawa}, {Bergin}, {Law}, {Loomis}, {Alarc{\'o}n}, {Andrews}, {Bae},
  {Bergner}, {Boehler}, {Booth}, {Bosman}, {Calahan}, {Cataldi}, {Cleeves},
  {Czekala}, {Furuya}, {Huang}, {Ilee}, {Kurtovic}, {Le Gal}, {Liu}, {Long},
  {M{\'e}nard}, {Nomura}, {P{\'e}rez}, {Qi}, {Schwarz}, {Sierra}, {Teague},
  {Tsukagoshi}, {Yamato}, {van't Hoff}, {Waggoner}, {Wilner}, \&
  {Zhang}}]{MAPS_oberg_2021ApJS..257....1O}
{{\"O}berg}, K.~I., {Guzm{\'a}n}, V.~V., {Walsh}, C., {et~al.} 2021, \apjs,
  257, 1, \dodoi{10.3847/1538-4365/ac1432}

\bibitem[{{Perez} {et~al.}(2015){Perez}, {Dunhill}, {Casassus}, {Roman},
  {Szul{\'a}gyi}, {Flores}, {Marino}, \&
  {Montesinos}}]{perez_2015ApJ...811L...5P}
{Perez}, S., {Dunhill}, A., {Casassus}, S., {et~al.} 2015, \apjl, 811, L5,
  \dodoi{10.1088/2041-8205/811/1/L5}

\bibitem[{Pinte {et~al.}(2022)Pinte, Teague, Flaherty, Hall, Facchini, \&
  Casassus}]{pinte_review_https://doi.org/10.48550/arxiv.2203.09528}
Pinte, C., Teague, R., Flaherty, K., {et~al.} 2022, Kinematic Structures in
  Planet-Forming Disks,  arXiv, \dodoi{10.48550/ARXIV.2203.09528}

\bibitem[{{Pinte} {et~al.}(2018{\natexlab{a}}){Pinte}, {Price}, {M{\'e}nard},
  {Duch{\^e}ne}, {Dent}, {Hill}, {de Gregorio-Monsalvo}, {Hales}, \&
  {Mentiplay}}]{pinte2018}
{Pinte}, C., {Price}, D.~J., {M{\'e}nard}, F., {et~al.} 2018{\natexlab{a}},
  \apjl, 860, L13, \dodoi{10.3847/2041-8213/aac6dc}

\bibitem[{{Pinte} {et~al.}(2018{\natexlab{b}}){Pinte}, {M{\'e}nard},
  {Duch{\^e}ne}, {Hill}, {Dent}, {Woitke}, {Maret}, {van der Plas}, {Hales},
  {Kamp}, {Thi}, {de Gregorio-Monsalvo}, {Rab}, {Quanz}, {Avenhaus}, {Carmona},
  \& {Casassus}}]{2018A&A...609A..47P}
{Pinte}, C., {M{\'e}nard}, F., {Duch{\^e}ne}, G., {et~al.} 2018{\natexlab{b}},
  \aap, 609, A47, \dodoi{10.1051/0004-6361/201731377}

\bibitem[{{Pinte} {et~al.}(2019){Pinte}, {van der Plas}, {M{\'e}nard}, {Price},
  {Christiaens}, {Hill}, {Mentiplay}, {Ginski}, {Choquet}, {Boehler},
  {Duch{\^e}ne}, {Perez}, \& {Casassus}}]{pinte_2019NatAs...3.1109P}
{Pinte}, C., {van der Plas}, G., {M{\'e}nard}, F., {et~al.} 2019, Nature
  Astronomy, 3, 1109, \dodoi{10.1038/s41550-019-0852-6}

\bibitem[{{Pinte} {et~al.}(2020){Pinte}, {Price}, {M{\'e}nard}, {Duch{\^e}ne},
  {Christiaens}, {Andrews}, {Huang}, {Hill}, {van der Plas}, {Perez}, {Isella},
  {Boehler}, {Dent}, {Mentiplay}, \& {Loomis}}]{2020ApJ...890L...9P}
{Pinte}, C., {Price}, D.~J., {M{\'e}nard}, F., {et~al.} 2020, \apjl, 890, L9,
  \dodoi{10.3847/2041-8213/ab6dda}

\bibitem[{{Sierra} {et~al.}(2021){Sierra}, {P{\'e}rez}, {Zhang}, {Law},
  {Guzm{\'a}n}, {Qi}, {Bosman}, {{\"O}berg}, {Andrews}, {Long}, {Teague},
  {Booth}, {Walsh}, {Wilner}, {M{\'e}nard}, {Cataldi}, {Czekala}, {Bae},
  {Huang}, {Bergner}, {Ilee}, {Benisty}, {Le Gal}, {Loomis}, {Tsukagoshi},
  {Liu}, {Yamato}, \& {Aikawa}}]{sierra2021}
{Sierra}, A., {P{\'e}rez}, L.~M., {Zhang}, K., {et~al.} 2021, \apjs, 257, 14,
  \dodoi{10.3847/1538-4365/ac1431}

\bibitem[{{Suriano} {et~al.}(2018){Suriano}, {Li}, {Krasnopolsky}, \&
  {Shang}}]{suriano2018}
{Suriano}, S.~S., {Li}, Z.-Y., {Krasnopolsky}, R., \& {Shang}, H. 2018, \mnras,
  477, 1239, \dodoi{10.1093/mnras/sty717}

\bibitem[{{Szul{\'a}gyi} {et~al.}(2014){Szul{\'a}gyi}, {Morbidelli}, {Crida},
  \& {Masset}}]{2014ApJ...782...65S}
{Szul{\'a}gyi}, J., {Morbidelli}, A., {Crida}, A., \& {Masset}, F. 2014, \apj,
  782, 65, \dodoi{10.1088/0004-637X/782/2/65}

\bibitem[{Teague(2019)}]{eddy}
Teague, R. 2019, The Journal of Open Source Software, 4, 1220,
  \dodoi{10.21105/joss.01220}

\bibitem[{{Teague} {et~al.}(2019{\natexlab{a}}){Teague}, {Bae}, \&
  {Bergin}}]{2019Natur.574..378T}
{Teague}, R., {Bae}, J., \& {Bergin}, E.~A. 2019{\natexlab{a}}, \nat, 574, 378,
  \dodoi{10.1038/s41586-019-1642-0}

\bibitem[{{Teague} {et~al.}(2018{\natexlab{a}}){Teague}, {Bae}, {Bergin},
  {Birnstiel}, \& {Foreman-Mackey}}]{teague_bae_2018ApJ...860L..12T}
{Teague}, R., {Bae}, J., {Bergin}, E.~A., {Birnstiel}, T., \& {Foreman-Mackey},
  D. 2018{\natexlab{a}}, \apjl, 860, L12, \dodoi{10.3847/2041-8213/aac6d7}

\bibitem[{{Teague} {et~al.}(2018{\natexlab{b}}){Teague}, {Bae}, {Birnstiel}, \&
  {Bergin}}]{as209_teague_2018ApJ...868..113T}
{Teague}, R., {Bae}, J., {Birnstiel}, T., \& {Bergin}, E.~A.
  2018{\natexlab{b}}, \apj, 868, 113, \dodoi{10.3847/1538-4357/aae836}

\bibitem[{{Teague} {et~al.}(2019{\natexlab{b}}){Teague}, {Bae}, {Huang}, \&
  {Bergin}}]{teague2019}
{Teague}, R., {Bae}, J., {Huang}, J., \& {Bergin}, E.~A. 2019{\natexlab{b}},
  \apjl, 884, L56, \dodoi{10.3847/2041-8213/ab4a83}

\bibitem[{{Teague} \&
  {Foreman-Mackey}(2018{\natexlab{a}})}]{2018RNAAS...2..173T}
{Teague}, R., \& {Foreman-Mackey}, D. 2018{\natexlab{a}}, Research Notes of the
  American Astronomical Society, 2, 173, \dodoi{10.3847/2515-5172/aae265}

\bibitem[{{Teague} \& {Foreman-Mackey}(2018{\natexlab{b}})}]{bettermoments}
---. 2018{\natexlab{b}}, {Bettermoments: A Robust Method To Measure Line
  Centroids}, v1.0,  Zenodo, \dodoi{10.5281/zenodo.1419754}

\bibitem[{Teague {et~al.}(2021)Teague, Law, Huang, \& Meng}]{disksurf}
Teague, R., Law, C.~J., Huang, J., \& Meng, F. 2021, Journal of Open Source
  Software, 6, 3827, \dodoi{10.21105/joss.03827}

\bibitem[{{Teague} {et~al.}(2021){Teague}, {Bae}, {Aikawa}, {Andrews},
  {Bergin}, {Bergner}, {Boehler}, {Booth}, {Bosman}, {Cataldi}, {Czekala},
  {Guzm{\'a}n}, {Huang}, {Ilee}, {Law}, {Le Gal}, {Long}, {Loomis},
  {M{\'e}nard}, {{\"O}berg}, {P{\'e}rez}, {Schwarz}, {Sierra}, {Walsh},
  {Wilner}, {Yamato}, \& {Zhang}}]{2021ApJS..257...18T}
{Teague}, R., {Bae}, J., {Aikawa}, Y., {et~al.} 2021, \apjs, 257, 18,
  \dodoi{10.3847/1538-4365/ac1438}

\bibitem[{{Teague} {et~al.}(2022){Teague}, {Bae}, {Andrews}, {Benisty},
  {Bergin}, {Facchini}, {Huang}, {Longarini}, \& {Wilner}}]{teague2022}
{Teague}, R., {Bae}, J., {Andrews}, S.~M., {et~al.} 2022, \apj, 936, 163,
  \dodoi{10.3847/1538-4357/ac88ca}

\bibitem[{{van der Walt} {et~al.}(2011){van der Walt}, {Colbert}, \&
  {Varoquaux}}]{numpy}
{van der Walt}, S., {Colbert}, S.~C., \& {Varoquaux}, G. 2011, Computing in
  Science and Engineering, 13, 22, \dodoi{10.1109/MCSE.2011.37}

\bibitem[{Virtanen {et~al.}(2020)Virtanen, Gommers, Oliphant, Haberland, Reddy,
  Cournapeau, Burovski, Peterson, Weckesser, Bright, {van der Walt}, Brett,
  Wilson, Millman, Mayorov, Nelson, Jones, Kern, Larson, Carey, Polat, Feng,
  Moore, {VanderPlas}, Laxalde, Perktold, Cimrman, Henriksen, Quintero, Harris,
  Archibald, Ribeiro, Pedregosa, {van Mulbregt}, \& {SciPy 1.0
  Contributors}}]{scipy}
Virtanen, P., Gommers, R., Oliphant, T.~E., {et~al.} 2020, Nature Methods, 17,
  261, \dodoi{10.1038/s41592-019-0686-2}

\bibitem[{{Wardle}(2007)}]{wardle2007}
{Wardle}, M. 2007, \apss, 311, 35, \dodoi{10.1007/s10509-007-9575-8}

\bibitem[{{W{\"o}lfer} {et~al.}(2022){W{\"o}lfer}, {Facchini}, {van der Marel},
  {van Dishoeck}, {Benisty}, {Bohn}, {Francis}, {Izquierdo}, \&
  {Teague}}]{wolfer_2022arXiv220809494W}
{W{\"o}lfer}, L., {Facchini}, S., {van der Marel}, N., {et~al.} 2022, arXiv
  e-prints, arXiv:2208.09494.
\newblock \doarXiv{2208.09494}

\bibitem[{Yu {et~al.}(2021)Yu, Teague, Bae, \& Öberg}]{Yu_2021}
Yu, H., Teague, R., Bae, J., \& Öberg, K. 2021, The Astrophysical Journal
  Letters, 920, L33, \dodoi{10.3847/2041-8213/ac283e}

\bibitem[{{Zhang} {et~al.}(2021){Zhang}, {Booth}, {Law}, {Bosman}, {Schwarz},
  {Bergin}, {{\"O}berg}, {Andrews}, {Guzm{\'a}n}, {Walsh}, {Qi}, {van't Hoff},
  {Long}, {Wilner}, {Huang}, {Czekala}, {Ilee}, {Cataldi}, {Bergner}, {Aikawa},
  {Teague}, {Bae}, {Loomis}, {Calahan}, {Alarc{\'o}n}, {M{\'e}nard}, {Le Gal},
  {Sierra}, {Yamato}, {Nomura}, {Tsukagoshi}, {P{\'e}rez}, {Trapman}, {Liu}, \&
  {Furuya}}]{zhang2021}
{Zhang}, K., {Booth}, A.~S., {Law}, C.~J., {et~al.} 2021, \apjs, 257, 5,
  \dodoi{10.3847/1538-4365/ac1580}

\bibitem[{{Zhang} {et~al.}(2018){Zhang}, {Zhu}, {Huang}, {Guzm{\'a}n},
  {Andrews}, {Birnstiel}, {Dullemond}, {Carpenter}, {Isella}, {P{\'e}rez},
  {Benisty}, {Wilner}, {Baruteau}, {Bai}, \&
  {Ricci}}]{zhang_2018ApJ...869L..47Z}
{Zhang}, S., {Zhu}, Z., {Huang}, J., {et~al.} 2018, \apjl, 869, L47,
  \dodoi{10.3847/2041-8213/aaf744}

\end{thebibliography}
\bibliographystyle{aasjournal}



\end{document}